\documentclass[10pt,a4paper,fleqn]{article}

\usepackage[
    top=20mm,
    bottom=20mm,
    left=18mm,
    right=18mm
]{geometry}

\usepackage{graphicx}
\usepackage{multicol}
\usepackage{titlesec}
\usepackage{fancyhdr}
\usepackage{setspace}
\usepackage{array}
\usepackage{booktabs}
\usepackage{tikz}
\usepackage{tcolorbox}
\usepackage{amssymb}
\usepackage{caption}
\usepackage{fontspec}
\usepackage{amsmath}
\usepackage{graphicx}
\usepackage{siunitx}
\usepackage{booktabs} 
\usepackage{float}
\usepackage{makecell}
\usepackage{multirow}
\usepackage{comment}

\usepackage{hyperref}
\usepackage{url}
\usepackage{lipsum}

\usepackage{booktabs}
\usepackage{titlesec}
\titleformat{\section}
    {\normalsize\bfseries}
    {\thesection}
    {1em}
    {}

\titleformat{\subsection}
  {\footnotesize\bfseries}
  {\thesubsection}
  {1em}
  {}

\titleformat{\subsubsection}
  {\footnotesize\bfseries}
  {\thesubsubsection}
  {1em}
  {}

\titleformat{\paragraph}[runin]
  {\footnotesize\bfseries}
  {\theparagraph}
  {1em}
  {}

\usepackage{lipsum}

\usepackage{varwidth}
\usepackage{adjustbox}

\definecolor{tether-gray}{HTML}{F2F1EF}

\newtcolorbox[auto counter]{custombox}[2][]{
    title={Box~\thetcbcounter: #2},
    label={#1},
    fonttitle=\bfseries,
    coltitle=black,
    before title=\vspace{15pt},
    colback=tether-gray,
    colframe=tether-gray,
    boxrule=0pt,
    arc=5pt,
    left=15pt,
    right=15pt,
    bottom=15pt,
    width=\textwidth,    
}

\newcommand{\tablebox}[1]{
    \tcbox[
        colback=tether-gray,
        colframe=tether-gray,
        boxrule=0pt,
        left=15pt,
        right=15pt,
        top=15pt,
        bottom=15pt
    ]{#1}
}

\begin{document}


\begin{flushleft}
\includegraphics[height=20pt]{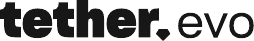}
\end{flushleft}

\vspace{10pt}


\begin{tcolorbox}[
    colback=tether-gray,
    colframe=tether-gray,
    boxrule=0pt,
    arc=5pt,
    left=15pt,
    right=15pt,
    top=15pt,
    bottom=15pt,
    width=\textwidth
]


{\huge\bfseries
Seeing the imagined: latent functional alignment in visual imagery decoding from fMRI data
}

\vspace{5pt}


{\small\bfseries
Fabrizio Spera\textsuperscript{1},
Tommaso Boccato\textsuperscript{2},
Michal Olak\textsuperscript{2},
Nicola Toschi\textsuperscript{1,3}\textsuperscript{*},
Matteo Ferrante\textsuperscript{1,2}\textsuperscript{*}
}

\vspace{2.5pt}

{\small
\textsuperscript{1}Department of Biomedicine and Prevention, University of Rome Tor Vergata, Rome, Italy\\
\textsuperscript{2}Tether Evo\\
\textsuperscript{3}A.A.
Martinos Center for Biomedical Imaging, Massachusetts General Hospital, Boston, USA\\
\textsuperscript{*} Equal Contribution
}

\vspace{15pt}


{\footnotesize
Recent progress in visual brain decoding from fMRI has been enabled by large-scale datasets such as the Natural Scenes Dataset (NSD) and powerful diffusion-based generative models. While current pipelines are primarily optimized for perception, their performance under mental-imagery remains less well understood. In this work, we study how a state-of-the-art (SOTA) perception decoder (DynaDiff) can be adapted to reconstruct imagined content from the NSD-Imagery benchmark. We propose a latent functional alignment (LFA) approach that maps imagery-evoked activity to the pretrained model's semantic content-enriched conditioning space, by adding a simple alignment module, while keeping the original remaining components frozen. To mitigate the limited amount of matched imagery–perception supervision, we further introduce a neural retrieval-based augmentation strategy that selects semantically related NSD perception trials from the same participants. Across four subjects, LFA consistently improves high-level semantic reconstruction metrics relative to the frozen pretrained baseline and a voxel-space ridge alignment baseline, and enables above-chance decoding from multiple cortical regions. These results suggest that semantic structure learned from perception can be leveraged to stabilize and improve visual imagery decoding under out-of-distribution conditions.
}

\end{tcolorbox}

\vspace{10pt}


\footnotesize


\section{Introduction}
\begin{figure}
\centering
\includegraphics[width=\linewidth]{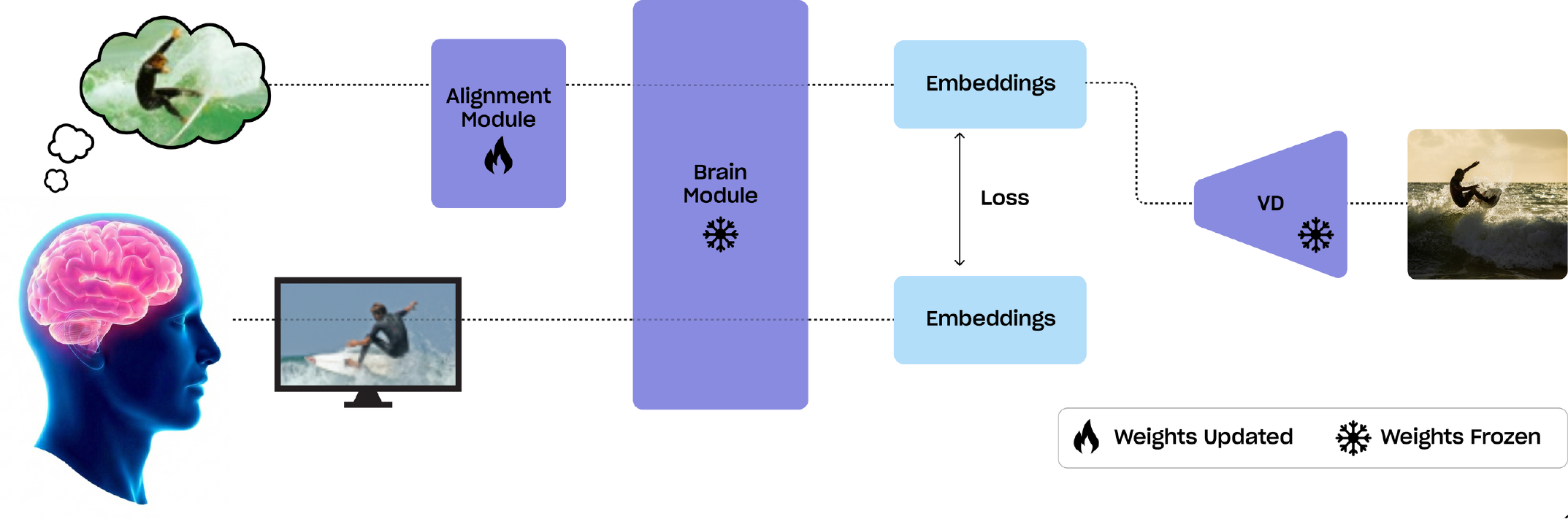}
\caption{\textbf{Overview of our pipeline}. CLIP-Image embeddings predicted by the pretrained visual decoding model from vision trials are used as targets for the alignment module. The module minimizes the loss between embeddings predicted from imagery trials and their corresponding visual targets, while all other components remain frozen.}
\label{pipe}
\end{figure}

Decoding the human brain is a rapidly evolving research area at the intersection of neuroscience and artificial intelligence (AI), with recent advances enabling the reconstruction of visual stimuli from neural activity and opening new perspectives for brain-computer interfaces (BCIs). Vision has received particular attention due to relatively controlled experimental settings and the increasing availability of large-scale datasets. Neural signals can be acquired using invasive techniques, such as electrocorticography (ECoG) and implanted microelectrodes, or non-invasive methods, including functional magnetic resonance imaging (fMRI), electroencephalography (EEG), and magnetoencephalography (MEG). Although invasive recordings generally provide higher signal-to-noise ratios (SNR), non-invasive techniques offer substantially lower risks. In particular, the release of the Natural Scenes Dataset (NSD) \cite{allen_massive_2022} demonstrated that competitive visual decoding can be achieved from fMRI using AI-based pipelines for image reconstruction.\\
Visual brain decoding typically maps noisy and high-dimensional brain data to natural images, often restricting the input to regions of interest (ROIs) in the visual cortex to reduce dimensionality. Neural activity can then be mapped into semantic embedding spaces learned by large pretrained models \cite{radford2021learningtransferablevisualmodels}, which provide conditioning signals for diffusion-based generative models.\\
In this work, we extend the SOTA visual perception decoding model, DynaDiff \cite{careil2025DynaDiffsinglestagedecodingimages}, to mental imagery. Mental imagery is a fundamental cognitive process through which individuals generate internal visual representations of past or future events. However, decoding imagery from brain activity is substantially more challenging than decoding visual perception, due to its lower SNR \cite{snr} and the limited availability of large-scale imagery datasets.\\
Recent results on the newly released NSD-Imagery benchmark dataset \cite{kneeland2025nsdimagerybenchmarkdatasetextending} show that models trained exclusively on visual perception data can produce moderately faithful reconstructions of imagined stimuli at inference time. However, existing open-source SOTA models exhibit inconsistent performance, likely due to the distribution shift between perception and imagery data and increased inter-subject variability. While semantic concepts may be shared across individuals, their internal visual representations can differ substantially, limiting the ability of existing pipelines to produce coherent reconstructions.\\
To address this distribution shift, we propose a strategy that encourages the model to exploit semantic representations shared between perception and imagery. We leverage the NSD-Imagery dataset together with the substantially larger NSD visual-perception dataset used to train the original model. Specifically, NSD visual-task neural data provide supervision targets for aligning the representations predicted from the imagery data, thereby leveraging the larger amount of available visual data to mitigate the limited size of the imagery dataset.\\
Finally, we systematically analyze the contribution of multiple cortical areas to assess whether the proposed alignment generalizes beyond the visual cortex. Our evaluation shows consistent improvements in semantic reconstruction metrics, demonstrating the effectiveness of the proposed approach for mental-imagery decoding. An overview of the proposed LFA pipeline is shown in Fig.~\ref{pipe}.

\subsection{Related Work}
Remarkable progress in visual brain decoding from fMRI data has been enabled by the availability of large-scale neural datasets \cite{god,bold5000,allen_massive_2022}, pretrained vision-language models such as CLIP \cite{radford2021learningtransferablevisualmodels}, and open-source diffusion models including Stable Diffusion (SD) \cite{rombach2022highresolutionimagesynthesislatent} and Versatile Diffusion (VD) \cite{xu2024versatilediffusiontextimages}. \\
Current SOTA visual brain decoding approaches typically map fMRI activity into high-dimensional CLIP-Image and/or CLIP-Text embedding spaces, which are subsequently used to condition pretrained diffusion models. Earlier approaches generally rely on multi-stage pipelines and linear models to reduce the temporal dimension of fMRI signals. Among these, MindEye1 \cite{scotti2023reconstructingmindseyefmritoimage}, BrainDiffuser \cite{ozcelik_natural_2023}, and MindEye2 \cite{scotti2024mindeye2sharedsubjectmodelsenable} constitute relevant baselines for visual decoding and have also been  evaluated in \cite{kneeland2025nsdimagerybenchmarkdatasetextending}.\\
MindEye1 employs separate semantic and perceptual pathways: fMRI activity is mapped to CLIP-Image embeddings for semantic reconstruction, while a second pathway predicts Stable Diffusion VAE \cite{kingma2022autoencodingvariationalbayes} latents to recover coarse perceptual structure and guide diffusion. BrainDiffuser similarly follows a two-stage framework, first predicting hierarchical VDVAE \cite{deepvae} latents for coarse image reconstruction and then refining the result with Versatile Diffusion conditioned on CLIP-Image and CLIP-Text representations predicted from fMRI. MindEye2 extends this paradigm through multi-subject pretraining followed by target-subject fine-tuning, while jointly modeling semantic and perceptual information and incorporating predicted image captions for textual guidance during reconstruction refinement.\\
More recently, DynaDiff \cite{careil2025DynaDiffsinglestagedecodingimages} introduced a single-stage visual decoding framework that directly exploits the temporal dynamics of fMRI signals. Its lightweight brain module applies subject-specific projections, timestep-dependent transformations, and temporal aggregation to map neural time series directly to embeddings matching the conditioning dimensionality of Versatile Diffusion. These embeddings replace CLIP-Image tokens in the U-Net cross-attention layers \cite{ronneberger2015unetconvolutionalnetworksbiomedical}, while the brain module and a LoRA adapter \cite{hu2021loralowrankadaptationlarge} are jointly optimized using the diffusion loss.\\
Despite its strong performance for visual perception, the ability of DynaDiff to generalize to mental imagery remains unexplored. We investigate this setting through LFA on NSD-Imagery.
\section{Methods}
\subsection{Data Augmentation}\label{augmentation}
\begin{figure}
\centering
\includegraphics[width=\linewidth]{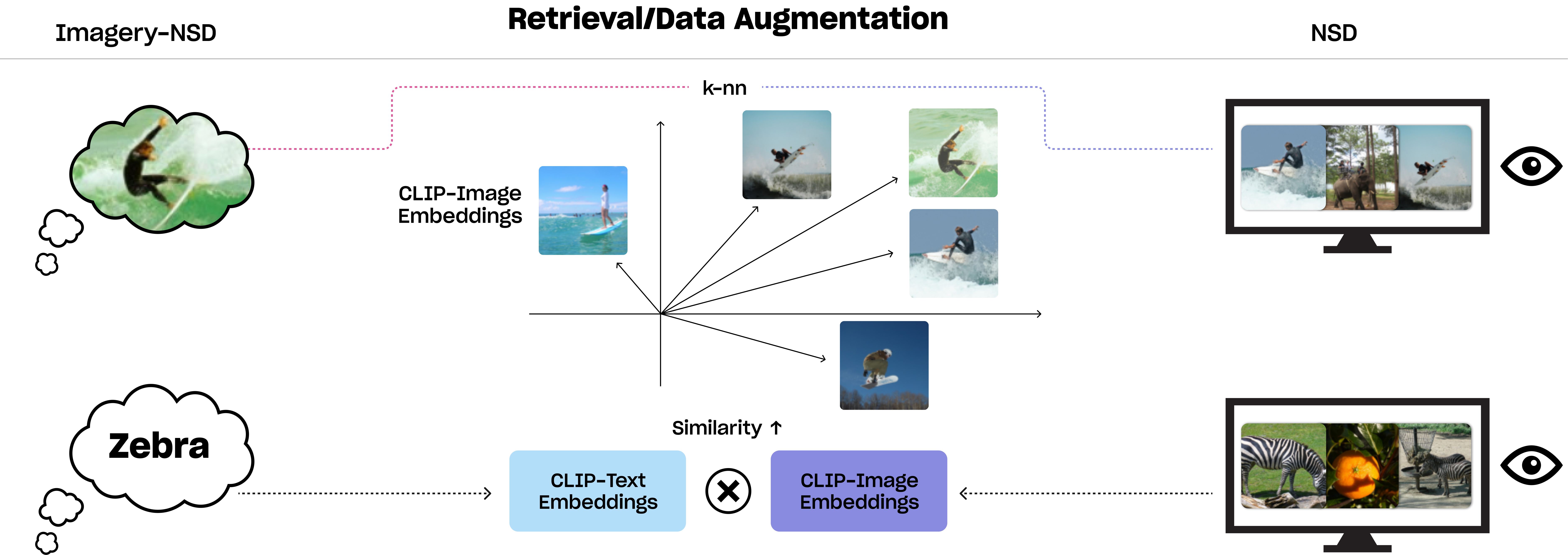}
\caption{\textbf{Illustration of the retrieval procedure}. 
For complex stimuli, we compute CLIP-Image embeddings and retrieve nearest neighbors from NSD via k-nn in cosine space, using the associated NSD brain data as targets. For conceptual stimuli, we compute CLIP-Text embeddings for the target words and retrieve NSD images via CLIP-Text/CLIP-Image similarity.}
\label{augment}
\end{figure}

One of the main challenges of the NSD-Imagery dataset is its limited amount of data available to construct a robust and efficient pipeline using only imagery data: each subject of the NSD performed additional dedicated runs on vision and imagination tasks on different categories of stimuli; these stimuli are divided into three different blocks, each one constituted of $6$ different stimuli, categorized as \textit{simple}, \textit{complex}, and \textit{conceptual} stimuli,  providing a total of $96$ imagery trials per block, and $48$ trials for the associated vision run; more details about the dataset can be found in Appendix\ref{app_dataset} and in \cite{kneeland2025nsdimagerybenchmarkdatasetextending}.\\ To overcome this difficulty, we retrieved additional fMRI data from NSD trials, focusing on complex and conceptual stimuli; in line with previous works, we focused only on subjects $1, 2, 5$ and $7$ since they performed a higher number of runs.\\
For each of the six complex stimuli we computed its CLIP-Image embedding and performed a nearest-neighbour search, in cosine distance, over the image embeddings of the NSD training split of the same subject (about 27,000 trials of 9,000 images, each shown three times). Five of the six complex stimuli belong to the NSD \textit{shared}$1000$ test-set, which is excluded from this split, and the sixth does not appear in NSD, so neither the target images nor their associated NSD trials can be retrieved. We retained $K$ nearest vision trials per stimulus and used them as visual targets. Reconstruction quality increased from $K = 0$ to $K = 1000$ as shown for subject $1$ in Appendix \ref{app_ablation}; therefore we used $K = 1000$ for all the four subjects.\\
Once we increased the number of visual data, each visual target had to be paired with an imagery input. Therefore, for each stimulus, the imagery training trials were replicated until their number matched that of the retrieved visual training targets, as explained in Appendix \ref{app_training}. To avoid exact duplicates, Gaussian noise with standard deviation $0.02\sigma$ was added to each copy beyond the original trials, where $\sigma$ is the standard deviation of the training trials for that stimulus. This replication was applied only to the training and validation data. A visual representation of our retrieval procedure from NSD is shown in Fig. \ref{augment}.
For conceptual stimuli, we embedded each target word with the CLIP-Text-encoder and retrieved the $K=1000$ NSD training trials whose image embeddings had the highest cosine similarity to it.\\
For simple stimuli, being outside the NSD distribution, no alignment module was trained  and imagery responses were decoded directly by the pretrained model.
\subsection{Latent Functional Alignment}\label{LFA}
A natural first approach is to learn, for each subject, a linear mapping from imagery to
perception responses directly in voxel space, analogous to the mappings used for
across-subject functional alignment \cite{ferrante_through_2024,art1,art2}. However, this approach aims to predict, in a very noisy and high-dimensional space, fine details that could not effectively capture the relationship between these two different cognitive processes. Therefore, our strategy relies on the hypothesis that mental images are typically less detailed and more variable than percepts, while their semantic content is more stable \cite{imagery_concepts,imagery1}: an alignment performed in the
embedding space that conditions the diffusion model could be more effective for this particular task, rather than capturing detailed differences in the voxel space. Following this ansatz, imagery responses are passed through an alignment module $f_\phi$ placed before the frozen
brain module $g$ of DynaDiff. The module first applies a learned linear mixing of the $T = 6$
time points, shared across voxels and initialised to the identity, since stimulus information
might not be uniformly distributed across time points; it then applies a multilayer perceptron with one hidden layer, with weights shared across time points, which maps the input voxels back to the $V_{\text{vis}}$ dimension expected by $g$. The Alignment module therefore operates in voxel space, but is
trained exclusively through a latent objective, while the original brain module and the diffusion model remain frozen.\\
Given an imagery input $\mathbf{x}$ paired with a retrieved perception response $\mathbf{v}$
of the same stimulus, we compare $\mathbf{e} = g(f_\phi(\mathbf{x}))$ with
$\mathbf{e}^{\text{vis}} = g(\mathbf{v})$, i.e. the embedding that the frozen brain module predicts from the perception response, minimizing
\begin{equation}
\mathcal{L} = \mathcal{L}_{\text{MSE}}\big(\mathbf{e}, \mathbf{e}^{\text{vis}}\big)
+ \lambda\, \mathcal{L}_{\text{SupCon}}\big(\mathbf{e}, \mathbf{e}^{\text{vis}}; \tau\big),
\qquad \lambda = 0.1,\ \tau = 0.05 .
\label{eq:loss}
\end{equation}
The $\mathcal{L}_{MSE}$ term regresses each input towards the mean target of its stimulus; with weakly informative imagery
inputs, predictions for different stimuli drift towards a common embedding. The supervised contrastive term \cite{contrastive} counteracts this collapse: each imagery embedding
acts as an anchor, the perception embeddings of the same stimulus in the batch act as positives
and those of the other stimuli as negatives, so that each prediction must be closer to the
targets of its own stimulus than to those of the others. More details on the training procedure can be found in Appendix \ref{app_training}.
\subsubsection{Evaluation}\label{evaluation}
In brain decoding there is no  single metric that fully captures the quality of reconstructed images. We therefore evaluate complementary aspects of reconstruction using low-level, high-level, and semantic metrics, motivated by the hierarchical correspondence between deep neural networks (DNNs) and the human visual system \cite{kriegeskorte_deep_2015,hierarchical2}. Low-level metrics, PixCorr and SSIM, measure pixel-wise correlation and structural similarity, respectively \cite{ssim}, providing an estimate of local visual fidelity but not of image meaning.
High-level metrics are computed from intermediate or final representations of pretrained DNNs and assess correspondence at increasingly abstract levels beyond pixel-level similarity. Specifically, AlexNet-2 and AlexNet-5 use layers 2 and 5 of AlexNet \cite{alex}, CLIP uses the output of the ViT-L/14 CLIP-Vision model \cite{radford2021learningtransferablevisualmodels}, and Inception uses the last pooling layer of InceptionV3 \cite{szegedy2015rethinkinginceptionarchitecturecomputer}. EffNet-B and SwAV are distance-based metrics derived from EfficientNet-B1 \cite{tan2020efficientnetrethinkingmodelscaling} and SwAV-ResNet50 \cite{caron2021unsupervisedlearningvisualfeatures}. For the identification-based metrics, each reconstructed image is compared with the correct target and with $11$ other distractors, following the evaluation protocol of \cite{kneeland2025nsdimagerybenchmarkdatasetextending}. The $11$ distractors are selected from the remaining six simple and six complex stimuli. Performance is the proportion of comparisons in which the correct target is more similar to the reconstruction than the distractor; therefore, chance level is $50\%$.
Since our approach aims to improve the semantic correspondence between imagery-based reconstructions and target stimuli rather than their low-level appearance, high-level metrics provide the most informative evaluation of our LFA.
We consider two baselines: first, we use the pretrained DynaDiff model, providing a reference for the generalization of the original visual decoding pipeline to imagery data. Second, as a baseline for functional alignment, we use cross-validated Ridge regression to predict visual-task fMRI activity from imagery-task fMRI activity in the voxel space.

\section{Results}\label{results}
\begin{figure}[H]
  \centering
  \includegraphics[width=\linewidth]{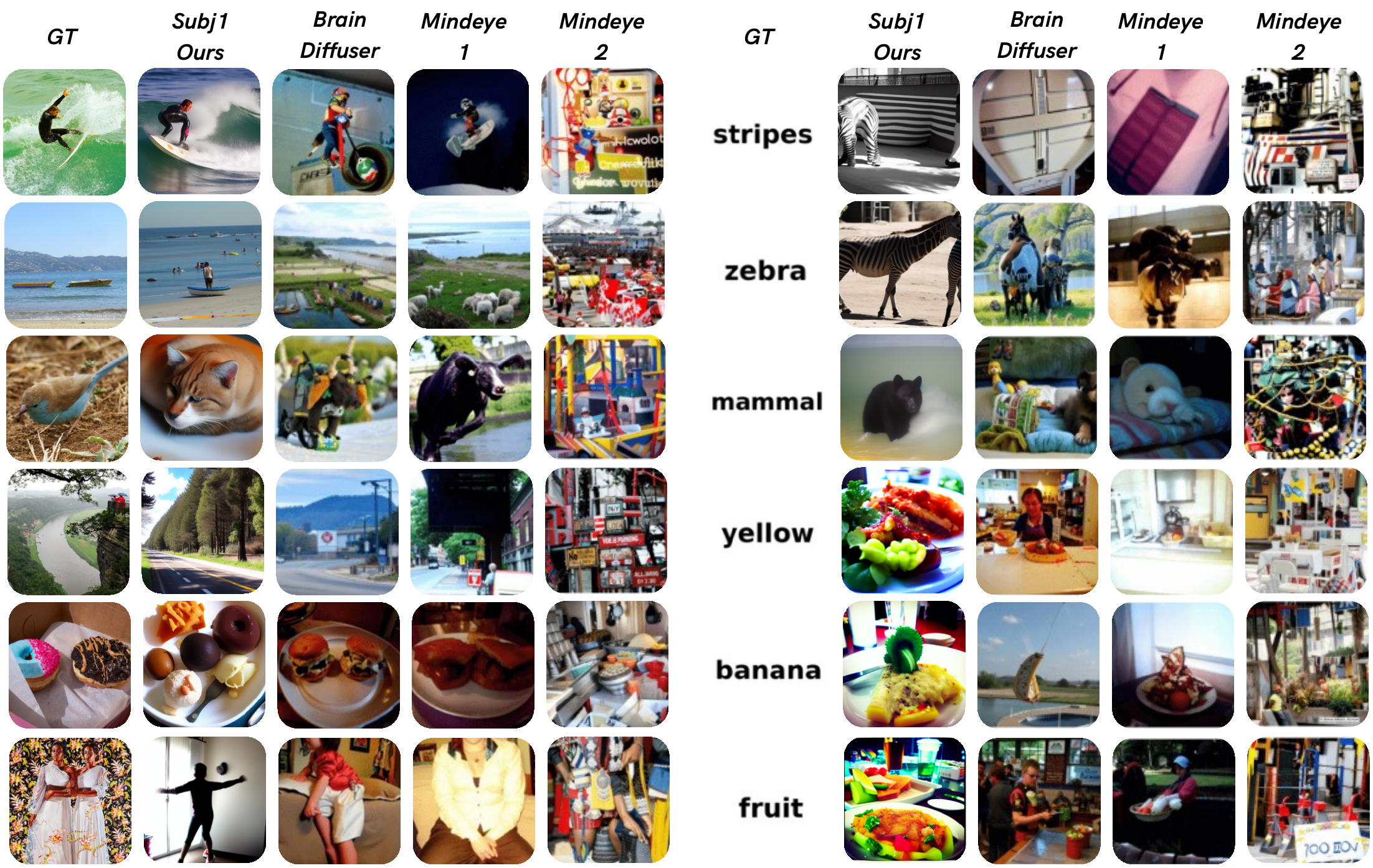}
  \caption{Qualitative results of the \textit{best} reconstruction for subject $1$ of imagined complex (left) and conceptual stimuli (right). Our results show a strong semantic consistency between the ground truth and reconstructed images, visually improving  the performance obtained by other SOTA models.}\label{rec_complex}
\end{figure}

In this section, we report the results of our LFA pipeline between
mental imagery and vision. As shown in Fig. \ref{rec_complex} for
subject $1$, our reconstructions are semantically coherent with the corresponding targets. The
reconstructions shown were selected by visual inspection as the best among 10 different 
generations, as done in
\cite{kneeland2025nsdimagerybenchmarkdatasetextending}; all quantitative results for single subjects are
averaged over all partitions, while  averaged subjects metrics are reported with standard error mean (SEM). For quantitative evaluation, no selection is performed: one reconstruction is generated for each of the 10 random partitions using a fixed sampling seed, and metrics are averaged across partitions.
Qualitatively, compared with the reconstructions of other open-source models
\cite{ozcelik_natural_2023,scotti2023reconstructingmindseyefmritoimage,scotti2024mindeye2sharedsubjectmodelsenable}
reported in \cite{kneeland2025nsdimagerybenchmarkdatasetextending}, our results appear more
photorealistic and semantically coherent. Even if fine details can be missed, the semantic
content is largely preserved, which is consistent with the hypothesis that mental images retain
their conceptual content more reliably than their low-level characteristics. Quantitative and qualitative reconstructions for
all subjects and stimulus classes, from the visual cortex, are
reported in Appendix \ref{app_quantitative} and Appendix \ref{app_qualitative}.
\subsection{Visual Brain Region}\label{visionsection}
\begin{figure}[h]
\centering
\includegraphics[width=\linewidth]{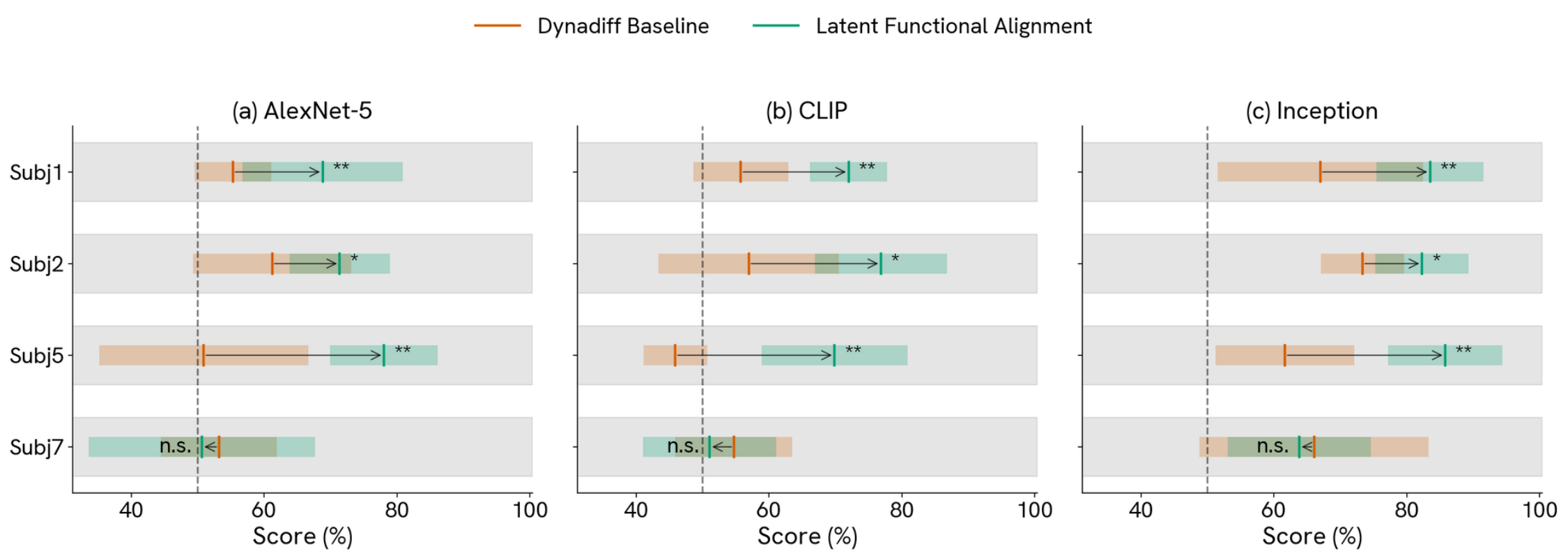}
\caption{Per-subject AlexNet-5, CLIP and Inception scores, with relative statistical significances. The gray vertical dashed line denotes chance level for each metric. DynaDiff baseline performs close to chance, whereas LFA improves performance across subjects, in particular we obtained significant improvements for subjects $1$, $2$ and $5$, while subject $7$ shows no significant improvement.}\label{clipealex}
\end{figure}

\begin{figure}[h]
\centering
\includegraphics[width=\linewidth]{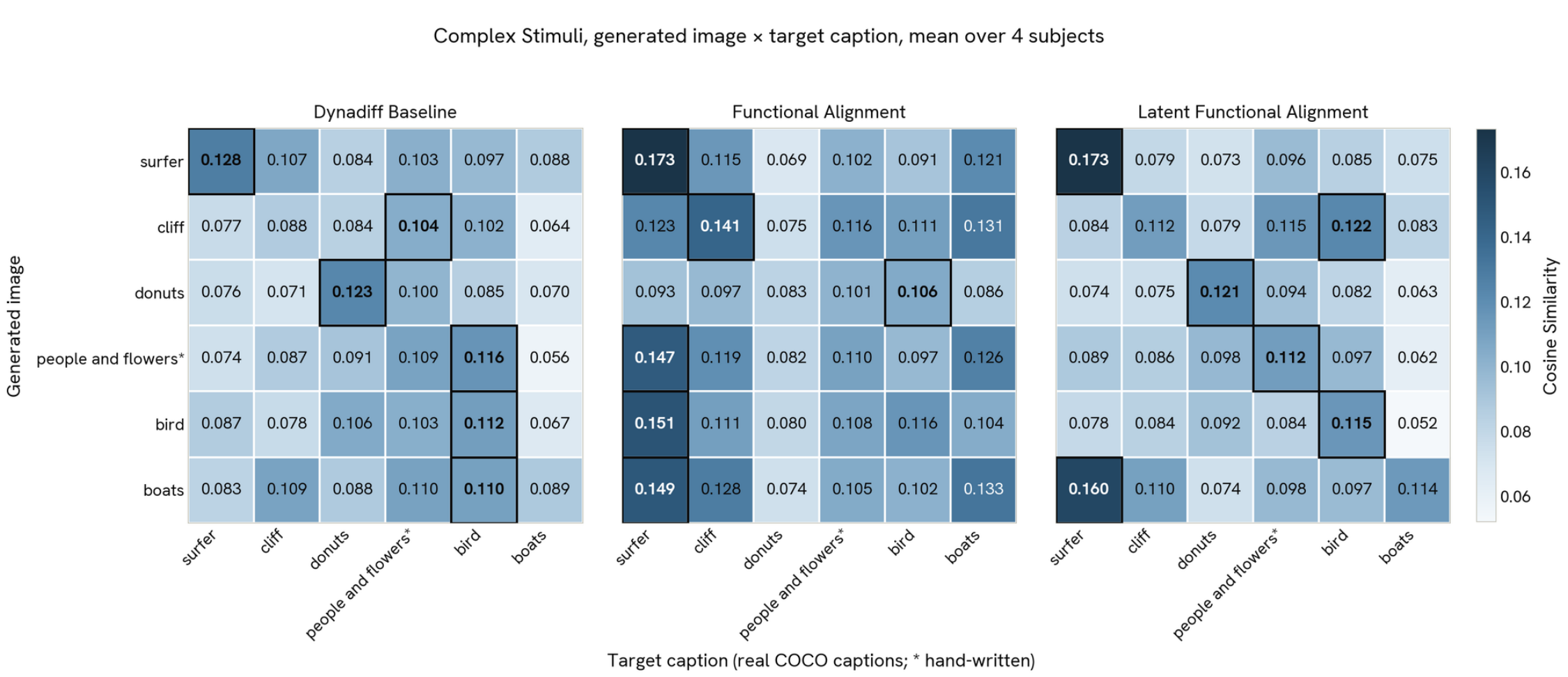}
\caption{Subject-averaged heatmaps for the three different methods reporting cosine similarity between CLIP-Text embeddings and CLIP-Image embeddings from reconstructions of complex stimuli. Functional Alignment, despite the  performance reported in Tab. \ref{vis_cortex}, shows significant overlapping, indicating a collapse in the generation process, especially for stimuli related to the presence of water, with only two terms along the diagonal indicating maximum correspondence. On the other hand we can see an improvement for LFA over the DynaDiff baseline along the diagonal, with four terms registering the maximum value for each row for our approach with respect to two maximum values for the baseline,  while preserving the separation among different concepts.}\label{heatmap_visivo} 
\end{figure}

We report in Tab. \ref{vis_cortex} the metrics averaged across the four subjects, computed only
for complex stimuli, since no alignment is applied to simple stimuli, whose reconstructions are
therefore identical to the baseline.
For each subject and partition, 20\% of the imagery trials
of each stimulus were held out and averaged into a single test input, in order to increase its SNR, yielding six
test reconstructions per stimulus set.
\begin{table}[h]
\centering
\small
\renewcommand{\arraystretch}{1.1}

\caption{Quantitative reconstruction metrics for complex stimuli averaged across four subjects; uncertainties are obtained with SEM. Arrows indicate whether higher ($\uparrow$) or lower ($\downarrow$) values correspond to better performance.}
\label{vis_cortex}
\tablebox{
\resizebox{0.92\textwidth}{!}{%
\begin{tabular}{lcccccccc}
\toprule
\textbf{Method} &
\multicolumn{2}{c}{\textbf{Low-Level}} &
\multicolumn{4}{c}{\textbf{High-Level}} &
\multicolumn{2}{c}{\textbf{Distance}} \\
\cmidrule(lr){2-3} \cmidrule(lr){4-7} \cmidrule(lr){8-9}
& PixCorr $\uparrow$ & SSIM $\uparrow$
& AlexNet-2 $\uparrow$ & AlexNet-5 $\uparrow$ & Incep $\uparrow$ & CLIP $\uparrow$
& Eff $\downarrow$ & SwAV $\downarrow$ \\
\midrule
\multicolumn{9}{c}{\textbf{NSD-Imagery Mental-Imagery Trials -- All Subjects -- Visual Cortex}} \\
\midrule
DynaDiff Baseline  &
$-0.004 \pm 0.010$ &
$0.241 \pm 0.003$ &
$76.86 \pm 1.64\%$ &
$55.15 \pm 2.21\%$ &
$67.01 \pm 2.41\%$ &
$53.33 \pm 2.52\%$ &
$0.961 \pm 0.004$ &
$0.622 \pm 0.003$ \\
Functional Alignment  &
$-0.006 \pm 0.010$ &
$0.189 \pm 0.011$ &
$77.05 \pm 1.21\%$ &
$\mathbf{68.90 \pm 3.25\%}$ &
$72.12 \pm 0.86\%$ &
$64.36 \pm 1.18\%$ &
$0.916 \pm 0.005$ &
$\mathbf{0.581 \pm 0.009}$ \\
Latent Functional Alignment  &
$\mathbf{0.035 \pm 0.007}$ &
$\mathbf{0.249 \pm 0.008}$ &
$\mathbf{80.98 \pm 4.23\%}$ &
$67.20 \pm 5.86\%$ &
$\mathbf{78.83 \pm 5.06\%}$ &
$\mathbf{67.42 \pm 5.65\%}$ &
$\mathbf{0.908 \pm 0.020}$ &
$0.583 \pm 0.017$ \\
\bottomrule
\end{tabular}%
}
}
\end{table}
With respect to the DynaDiff baseline, LFA improves all high-level and
distance metrics, as well as PixCorr. Functional alignment also improves several high-level
metrics, and outperforms our method on AlexNet-5 and SwAV, but at the cost of a marked decrease
in SSIM. \\ Fig. \ref{clipealex} shows AlexNet-5, CLIP and Inception scores before and after LFA for each subject; significance was assessed with a one-sided Wilcoxon
signed-rank test over the 10 partitions.
From Tab. \ref{vis_cortex}, Functional Alignment obtains strong high-level scores, but these scores can be inflated by representations that capture shared semantic structure across stimuli . We therefore investigate stimulus specificity separately, computing the similarity between CLIP-Image embeddings of the generated images and the CLIP-Text embeddings from the COCO captions \cite{lin2015microsoftcococommonobjects} of each of the 5 stimuli from the NSD \textit{shared}$1000$ set, while the remaining complex stimulus caption has been hand-written. Results are reported in Fig. \ref{heatmap_visivo}. 
For conceptual stimuli, visual targets were retrieved from NSD as described in
Section \ref{augmentation}. Since no ground-truth image exists, we
generated a caption for each reconstruction with BLIP \cite{BLIP} and computed the cosine
similarity between the MPNet \cite{MPNET} sentence embeddings of the captions and of the target
words. Results are shown in Tab. \ref{tab_concept_similarity}.
\begin{table}[H]
\centering
\small
\renewcommand{\arraystretch}{1.1}

\caption{Cosine similarity between BLIP-generated captions and target words for conceptual stimuli in MPNet sentence embeddings space. Values are reported as mean $\pm$ standard deviation across the ten reconstruction trials for each subject and as mean $\pm$ SEM across subjects.}
\label{tab_concept_similarity}
\tablebox{
\resizebox{0.92\textwidth}{!}{%
\begin{tabular}{lccccc}
\toprule
\textbf{Method} &
\textbf{Subj1} &
\textbf{Subj2} &
\textbf{Subj5} &
\textbf{Subj7} &
\textbf{All Subjects (Mean $\pm$ SEM)} \\
\midrule
\multicolumn{6}{c}{\textbf{NSD-Imagery Conceptual Stimuli -- Text Similarity -- Visual Cortex}} \\
\midrule
DynaDiff Baseline &
$0.109 \pm 0.050$ &
$0.119 \pm 0.048$ &
$0.082 \pm 0.036$ &
$0.086 \pm 0.028$ &
$0.099 \pm 0.009$ \\
Functional Alignment &
$0.184 \pm 0.040$ &
$\mathbf{0.223 \pm 0.044}$ &
$0.183 \pm 0.034$ &
$\mathbf{0.163 \pm 0.053}$ &
$0.188 \pm 0.012$ \\
Latent Functional Alignment &
$\mathbf{0.187 \pm 0.027}$ &
$0.156 \pm 0.041$ &
$\mathbf{0.190 \pm 0.042}$ &
$0.152 \pm 0.035$ &
$0.171 \pm 0.010$ \\
\bottomrule
\end{tabular}%
}
}
\end{table}
As for complex stimuli, functional alignment reaches a high similarity with the target words,
even slightly higher than LFA. However,
Tab. \ref{tab_concept_collapse} shows that this similarity is not specific to the target: since
the six words form two semantic groups (\textit{banana}, \textit{yellow}, \textit{fruit}; \textit{zebra}, \textit{stripes}, \textit{mammal}),
a generic reconstruction can match several words at once. 
\begin{table}[H]
\centering
\small
\renewcommand{\arraystretch}{1.1}

\caption{Difference between the mean cosine similarity of diagonal and non-diagonal terms, weighted by the inverse number of corresponding terms, in the text-similarity matrix for the DynaDiff baseline, Functional Alignment, and LFA on conceptual stimuli.}
\label{tab_concept_collapse}
\tablebox{
\resizebox{0.92\textwidth}{!}{%
\begin{tabular}{lccccc}
\toprule
\textbf{Method} &
\textbf{Subj1} &
\textbf{Subj2} &
\textbf{Subj5} &
\textbf{Subj7} &
\textbf{All Subjects (Mean $\pm$ SEM)} \\
\midrule
\multicolumn{6}{c}{\textbf{NSD-Imagery Conceptual Stimuli -- Specificity -- Visual Cortex}} \\
\midrule
DynaDiff Baseline &
$0.039 \pm 0.043$ &
$0.050 \pm 0.031$ &
$0.007 \pm 0.033$ &
$0.014 \pm 0.030$ &
$0.028 \pm 0.010$ \\
Functional Alignment &
$0.035 \pm 0.038$ &
$0.074 \pm 0.045$ &
$0.038 \pm 0.033$ &
$0.012 \pm 0.052$ &
$0.040 \pm 0.013$ \\
Latent Functional Alignment &
$\mathbf{0.073 \pm 0.026}$ &
$\mathbf{0.075 \pm 0.031}$ &
$\mathbf{0.079 \pm 0.040}$ &
$\mathbf{0.040 \pm 0.047}$ &
$\mathbf{0.067 \pm 0.009}$ \\
\bottomrule
\end{tabular}%
}
}
\end{table}
\subsection{Role of non-visual brain regions}\label{other_rois}
\begin{figure}[h]
\centering
\includegraphics[width=\linewidth]{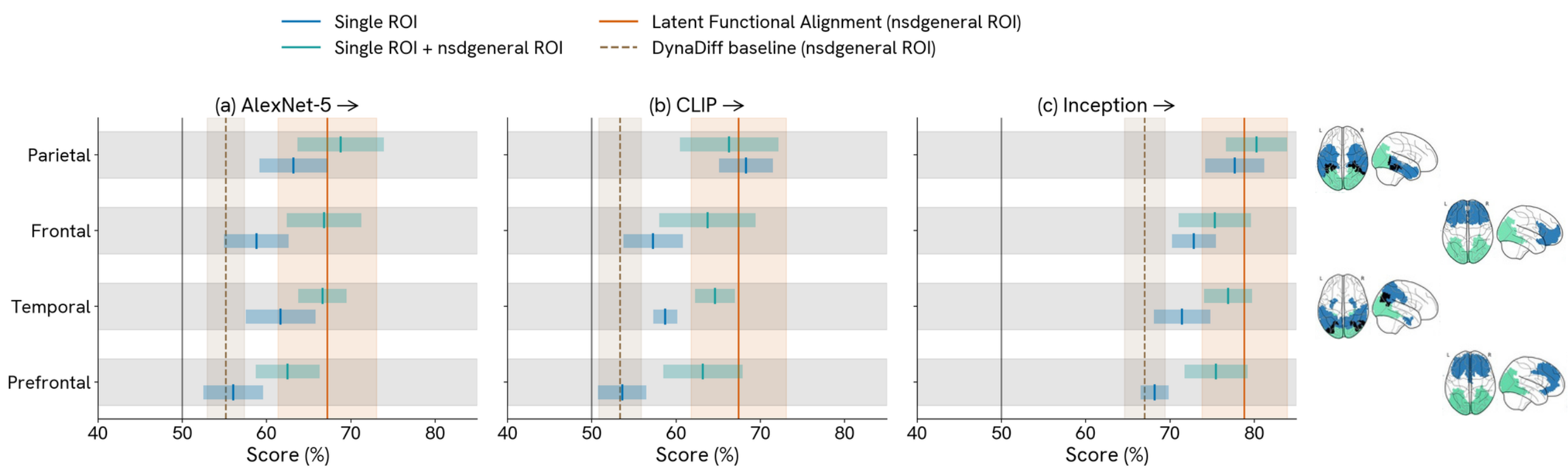}%
  \caption{Bar plot averaged across the four subjects for AlexNet-5, CLIP and Inception for complex stimuli reconstructions. Results are obtained from the single ROI voxels (blue bars) and their combination with the \textit{nsdgeneral} ROI (greenish bars). The comparison is performed both on DynaDiff Baseline (vertical brown dashed line) and LFA (vertical orange line) from the \textit{nsdgeneral} ROI, taking the values reported in Tab. \ref{vis_cortex}.
  On the right side there are the corresponding regions activated; the teal color refers to the posterior cortex, while blue to the other ROIs; the black points indicate the overlap between the posterior cortex and the considered ROI.  }\label{mean_rois}
\end{figure}

\begin{figure}[h]
\centering
\includegraphics[width=\linewidth]{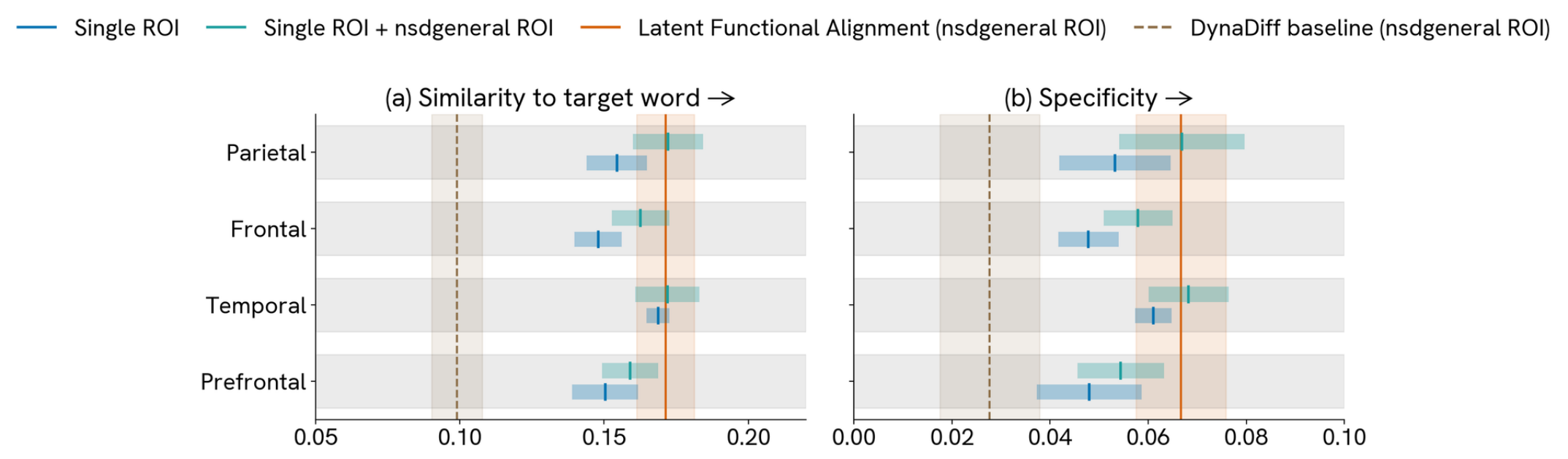}%
  \caption{Panel (a): Similarity between generated Image embeddings and Text embeddings for conceptual stimuli for different ROIs. Both values related to single areas and with the addition of the \textit{nsdgeneral} ROI are above DynaDiff baseline (vertical brown dashed line). Panel (b): difference between the mean of the diagonal and non-diagonal terms, weighted by the inverse of their respective number of terms, for each ROI and its combination with the \textit{nsdgeneral} ROI.
  In all the cases the addition of visual cortex improve performance, in particular for parietal and temporal regions they tend to the same values of the LFA (vertical orange dashed line) results in Tab. \ref{tab_concept_similarity}. }\label{mean_rois_concepts}
\end{figure}

The mechanisms and the cortical regions involved in mental imagery are still debated
\cite{imagery2}, and the role of visual cortex in particular has yielded mixed
findings across studies and individuals
\cite{imagery3,imagery4,imagery5,imagery6,vis-imagery1,vis-imagery2}. Here we compare
quantitatively the contribution of the main cortical areas, for both complex and conceptual
stimuli.\\
We selected each ROI from the HCP$-$MMP$1$ atlas in the functional space of each subject, with the
same resolution and preprocessing used for the \textit{nsdgeneral} ROI (\ref{app_dataset}), and repeated the
procedure described in Section \ref{visionsection} using the voxels as input, indicated in Tab. \ref{rois_voxels} in Appendix \ref{app_training}. We then repeated it on the
set union of each ROI with \textit{nsdgeneral}, to measure the interplay between the posterior
cortex and each region. Note that no unaligned baseline exists over \textit{nsdgeneral} ROI, since the pretrained brain module only accepts \textit{nsdgeneral} voxels: every
value reported here is obtained with LFA, and the baseline shown for
reference is that of the visual cortex from Section \ref{visionsection}. The results averaged across the four subjects are
reported in Fig. \ref{mean_rois}. 
For complex stimuli, every region, except the prefrontal cortex, improves over the visual-cortex
baseline, with the parietal cortex as the strongest single region. Adding the \textit{nsdgeneral} ROI to each region brings all of them close to the
values obtained from the visual cortex alone, and never above it; the only region whose performance is not improved by this
addition is the parietal cortex on CLIP. 
Overall, this indicates that the information used for reconstructing imagined complex stimuli
comes mainly from the posterior cortex, as pointed out from previous neuroscientific works \cite{visione-immaginazione2,imagery_concepts,immaginazione-visione1}.\\
We investigated possible contributions distributed across the cortex also for conceptual stimuli; in Fig. \ref{mean_rois_concepts} we reported caption-to-word similarity in MPNet
space described in Section \ref{visionsection}.
Here the picture differs: every single region is closer to LFA from the
visual cortex  than to the baseline, with the temporal lobes the closest. The contribution of \textit{nsdgeneral}
increases performance for all regions, but there is still no combination that exceeds the visual cortex alone. A
possible interpretation is that conceptual imagery relies on a more distributed and less
posterior representation than the imagery of specific complex images, but more experiments are needed to confirm this. Additional details on statistical reliability of our analysis can be found in Appendix \ref{stats}.

\section{Discussion}
In this work, we investigated brain decoding from visual-imagery by analyzing the
contribution of distinct cortical regions. Compared to visual perception, imagery is a more challenging decoding target, as it is driven by internally generated
working-memory and top-down processes rather than by direct sensory input \cite{top-down1,top-down2}.\\
As an initial step, we focused on visual cortical regions, leveraging a pretrained perception-model
\cite{careil2025DynaDiffsinglestagedecodingimages}, following the procedure of
\cite{kneeland2025nsdimagerybenchmarkdatasetextending} and previous research attributing to visual cortex a major role in visual-imagery \cite{vis-imagery1,vis-imagery2}. The proposed LFA framework
extracts semantically meaningful representations from imagery-related signals: averaged across four participants, it improves every high-level and distance metric over the unaligned
model, and the improvement is significant in three subjects out of four. Subject $7$ is the
exception, showing no improvement on any high-level metric, which is consistent with the weak
and variable nature of imagery signals and with the inter-subject variability reported in
previous work \cite{snr,sub_variaibility}.\\
The comparison with voxel-space functional alignment also carries a methodological message.
That approach was competitive on the standard reconstruction
metrics, while producing reconstructions that were not specific to the imagined stimulus. Only
measuring the match to the correct target with respect to the others
revealed the difference. Therefore, metrics designed for perceptual decoding should not be used alone to assess imagery reconstructions, as they may overlook potential discrepancies with human judgments \cite{metrica1}.\\
Subsequently, we extended the analysis to the parietal, temporal, frontal, and prefrontal regions.
Posterior cortex emerged as the primary contributor to decoding performance, as observed by \cite{wave}: for complex stimuli, reconstructions based on individual non-visual ROIs yielded
lower scores than the visual cortex, although all regions, except the prefrontal cortex, are above the visual-cortex baseline, with the parietal cortex the strongest single region. Combining each ROI with the \textit{nsdgeneral} ROI improved the results of the single non-visual ROI, but never exceeded those obtained from the \textit{nsdgeneral} ROI by itself. One plausible explanation is that, given a fixed visual alignment target, a single additional ROI does not contribute sufficiently complementary information, an effect likely amplified by the limited amount of imagery data
available to learn cross-regional interactions.\\
For conceptual stimuli the picture differs: every single region, prefrontal cortex included, was closer to the values obtained from the visual cortex than to the baseline, with the temporal
lobes the closest. This suggests that the representation supporting conceptual imagery is less confined to posterior cortex than the one supporting the imagery of specific images, although
our data-driven approach cannot establish this directly.\\
More broadly, these results are compatible with the idea that  although internally generated representations differ fundamentally from perceptual ones, they remain semantically related, and the ability of the model to exploit them underscores its sensitivity to non-sensory neural
representations. Overall, our findings support the view of visual imagery as a distributed process engaging both sensory and higher-order cognitive regions.
\subsection{Limitations and Conclusion}
Several limitations must be acknowledged. First, the limited availability of imagery data, combined with the strong inter-subject variability inherent to mental imagery, restricts the
neuroscientific conclusions that can be drawn about the functional role of individual ROIs; the absence of any improvement in subject $7$ is the clearest illustration. Second, the alignment module is trained and evaluated on the same six stimuli per set, with held-out trials, so our results demonstrate generalization across trials and not to unseen stimuli; the correlation analysis in Appendix \ref{app_dataset} further characterizes this setting, showing substantial variability across imagery trials and providing additional context for interpreting this trial-level generalization. Third, no unaligned
baseline exists for non-visual regions, since the pretrained brain module accepts only \textit{nsdgeneral} voxels: ROI results can be compared with each other and with the visual cortex, but
not with an unaligned counterpart of the same region. Fourth, the parietal and temporal ROIs partially overlap with the \textit{nsdgeneral} ROI, so they are not purely non-visual; the frontal and
prefrontal ROIs, which do not overlap, provide the cleanest evidence that imagery-related information extends beyond posterior cortex.\\
Finally, while retrieval-based augmentation improved performance substantially, relying on perception data may bias the alignment towards perceptual
rather than purely imagery-driven features, and for conceptual stimuli the retrieved target sets of semantically related words partially overlap, which limits the separation the model can
learn.\\
Despite these limitations, our results indicate that visual imagery engages a broad and distributed cortical network, supporting multi-ROI decoding as a promising direction for BCIs and semantic reconstruction from neural signals, with evidence on the primary role played by the posterior area for visual-imagery. Future work will explore more sophisticated ROI fusion, finer parcellations, imagery-specific evaluation metrics, and validation across larger cohorts and more diverse semantic domains. Ultimately, understanding how imagination is encoded in the brain may not only improve decoding performance but also deepen our understanding of the neural foundations of creativity and inner experience. Such knowledge could also pave the way for future clinical applications, particularly for patients who are unable to communicate or see but retain the ability to internally evoke previously experienced scenes and concepts.

\subsection*{AI use statement}
In this work, we used generative AI tools for polishing text in English.
We have not used generative AI tools for generating concepts and procedure ideas. We have reviewed all AI-assisted work. Part of the code was LLM-generated and was verified and tested for correctness
by the authors. We take responsibility for the final content of this work,
including text, claims or artifacts produced with the aid of generative AI.

\subsection*{Ethics statement}
The datasets are publicly accessible at: \url{https://natural-scenes-dataset.s3.amazonaws.com/index.html}. The present study
makes use of the publicly available fMRI datasets \cite{allen_massive_2022,kneeland2025nsdimagerybenchmarkdatasetextending}, which were collected under informed consent
and institutional ethical approval. The authors did not acquire new data. All data handling adheres to the licenses and usage terms specified by the dataset providers, and no attempt was made to re-identify individuals or to infer private attributes unrelated to the stated research purposes. All the data are acquired from healthy subjects who do not suffer from aphantasia, or related clinical problems. 
Results and research efforts in decoding human mind raise broader conceptual and ethical questions. As neural data become more accessible and AI models more powerful, the information that can be decoded from brain activity continues to expand. Decoding imagination, one of the most private aspects of human experience, offers new opportunities for human--AI co-creation and the externalization of memories, dreams, or creative intent, while raising concerns about privacy, neural rights, model bias, and interpretability. Key open questions include how to safeguard individuals against misuse and how to disentangle neural noise, subjective variability, and algorithmic bias in decoded representations.\\

\subsection*{Reproducibility statement}
To reproduce our results, we recommend following the pipeline in the following order. First, preprocess the NSD-Imagery dataset as described in Appendix \ref{app_dataset}, to ensure compatibility with the pretrained perception model. Second, apply the data augmentation pipeline described in Section \ref{augmentation} to obtain the data required by the LFA method, whose training procedure is detailed in Appendix \ref{app_training} and Section \ref{LFA}. In order to include the other ROIs use the same parcellation of Section \ref{other_rois}, from HCP$-$MMP$1$ atlas.\\
Finally, follow the evaluation protocol described in Section \ref{evaluation} to reproduce the results reported in Section \ref{results}.

\bibliographystyle{ieeetr}
\bibliography{bib}

\appendix
\section{Appendix}

\subsection{Dataset and Preprocessing}\label{app_dataset}
\begin{figure}[H]
\small
\centering
\includegraphics[width=\linewidth]{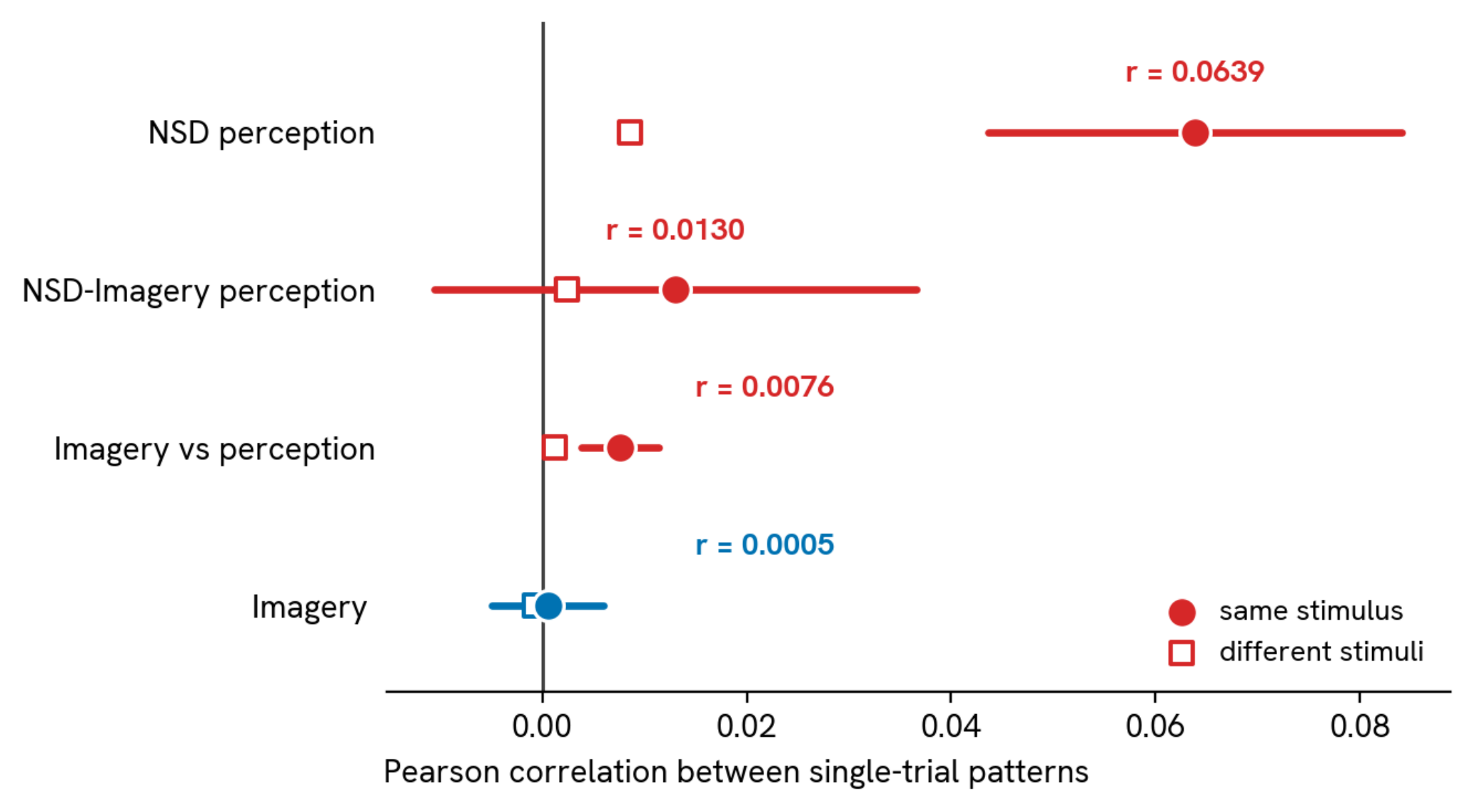}
\caption{Pearson correlation between single trial fMRI patterns (voxels × time samples). Filled circles: pairs of trials of the same stimulus, mean and 95$\%$ confidence interval across blocks; open squares: pairs of different stimuli. The lower three rows average four subjects on the three different stimuli sets; the top row averages NSD perceptual repetitions over the four subjects. Two imagery trials of the same stimulus are uncorrelated ($r = 0.0005$), whereas two perceptual repetitions of the same NSD image reach $r = 0.0639$.}
\label{variability}
\end{figure}

The NSD provides high resolution fMRI measurements from eight participants, each exposed to $9000-10000$ distinct natural color images over $30–40$ scanning sessions at $7$T. The stimuli were drawn from the Microsoft Common Objects in Context (COCO) dataset \cite{lin2015microsoftcococommonobjects}, with a shared subset of $1000$ images and subject specific unique images. All participants provided written informed consent prior to the $7$T fMRI sessions, including explicit authorization for anonymized data sharing with the research community; the study protocols were approved by the Institutional Review Board (IRB) at the University of Minnesota in accordance with ethical standards for human subjects research.\\
The NSD-Imagery dataset includes $12$ additional runs per NSD participant divided into different stimulus and task classes. The dataset contains $18$ stimuli in total, divided into six simple stimuli (single or double bars at various orientations), six complex stimuli (five of them taken from the NSD \textit{shared}$1000$ test set with one novel image), and six conceptual stimuli (target words). Each stimulus is repeated eight times per run, with $48$ trials per run and it has been associated with a particular cue letter that the participants memorized in practice sessions before scanning sessions.\\
The runs consist of vision tasks for each stimulus type ($3$ runs) with the presentation of both stimulus and a cue letter, that may or may not match the corresponding stimulus on the screen, and imagination tasks, repeated twice per stimulus ($6$ runs); in these runs, only cue letters were presented on the screen to the participants, who were instructed to use the cue letter to mentally visualize the associated stimulus. Three additional runs referred to attention tasks were not related to imagination and we excluded them from our study. We used the data from the same subjects we considered in the NSD.\\
In our analyses, visual task runs from both the NSD and NSD-Imagery datasets analyze the brain activity of each subject from the \textit{nsdgeneral} ROI, manually drawn on fsaverage and encompassing voxels responsive to the posterior cortex \cite{allen_massive_2022}. Additional ROIs used exclusively for the imagination task were defined by parcellating the cortex according to the HCP$-$MMP$1$ atlas \cite{glasser_multi-modal_2016}.
Since the DynaDiff model captures the temporal dynamics of the neural signals, we worked directly with the time series data; keeping the same prescriptions used for the pretrained model, we selected data from the functional space of the subjects with an isotropic resolution of $1.8$ mm; additionally, a cosine-drift linear model to each voxel in the time series was subtracted from the raw signal; finally, each voxel time series is z-score normalized after each run, in order to not compromise the specificity of the data, linked to the type of task performed, each with its own SNR, as suggested by \cite{kneeland2025nsdimagerybenchmarkdatasetextending}.
Since the time resolution (TR) of acquisition of the NSD-Imagery is equal to $1.0$ seconds, but the pretrained model expects data with TRs of $1.33$ seconds, we performed per-voxel linear interpolation in order to resample them as expected from the model.
As reported in the NSD-Imagery benchmark, approximately $50\%$ of vision task trials exhibit a mismatch between the displayed image and the corresponding cue letter. We therefore discarded these trials from our analysis. Consequently, half of the $48$ total vision trials per run remained usable, making any alignment procedure relying only on the NSD-Imagery data-limited and challenging to implement.\\
Before conducting experiments, we verified results on vision runs from NSD-Imagery for all the $4$ subjects, reporting in Tab. \ref{tab:vision_mean} in Appendix \ref{app_quantitative} averaged-subjects metrics with SEM and qualitative results in Fig. \ref{all_vision} in Appendix \ref{app_qualitative}.
As expected, reconstruction quality improves when using visual perception data rather than visual imagery data, further validating the preprocessing procedure.\\
We conducted additional analyses of the correlations between brain activity patterns across trials for both the NSD and NSD-Imagery datasets, considering both visual perception and visual imagery, respectively, across all four subjects. Fig. \ref{variability} reports the Pearson correlation coefficients estimated from the neural data, revealing differences both across modalities and between datasets.
The first and second rows report the correlations for the visual perception task in the NSD and NSD-Imagery datasets, respectively. As also reported in \cite{kneeland2025nsdimagerybenchmarkdatasetextending}, performance on NSD-Imagery is lower than on NSD, possibly due to differences in the experimental settings of the two datasets.
The third row shows the correlation between visual imagery and visual perception responses in the NSD-Imagery runs; although relatively small, a measurable correlation is present, providing further support for aligning visual-imagery representations with those learned from visual perception data. Finally, the last row reports the Pearson correlations between imagery trials. This is the only case in which we observe a partial overlap between responses to different trials involving the same stimulus and trials involving different stimuli. This pattern is consistent with the strong variability and near-chance performance observed with the DynaDiff baseline, while also supporting the generalization of our approach across trials, given the very weak correlation between different imagery trials.

\subsection{Details on training procedure}\label{app_training}

During the training of the alignment module we used an AdamW optimizer with a learning rate of $10^{-3}$ and weight decay $10^{-2}$, with a batch size of $36$; all the other modules of the original architecture were kept frozen. Training ran for at most $30$ epochs with early stopping after $8$ epochs without improvement, and the retained model was the one maximising the difference between the diagonal and the off-diagonal entries of the matrix of cosine similarities between predicted and target CLIP-Image embeddings, on the validation set. The diagonal and off-diagonal terms are weighted according to the number of entries in each group, with weights equal to the inverse of their respective counts. The imagery data were split by trial, reserving $20\%$ for test and $15\%$ of the remainder for validation; the augmented targets were split independently, $90\%$ for training and $10\%$ for validation, keeping the groups of three repetitions of the same NSD image such that no repetition is shared between the two sets.
The alignment module, both for the visual cortex and for the individual ROIs, consists of the temporal mixing stage followed by a multilayer perceptron with a single hidden layer of $8000$ units, layer normalization, \textit{GELU} activation and dropout with probability $0.25$. Its output dimensionality matches the voxel count expected by the original brain module, which corresponds to the number of voxels, $V_{vis}$, in the \textit{nsdgeneral} ROI for each subject.
Formally, we indicate the input time series data as $\mathbf{X}$, such that $\mathbf{X} \in \mathbb{R}^{B \times V \times T}$, where $B$ is the batch dimension, $V$ is the number of input voxels that depends both on the subject and the ROI considered as reported in Tab. \ref{rois_voxels}, while $T$ is the number of time points, with $T=6$.\\
The first transformation is a temporal mixing stage, introduced because the subsequent modules process the $T$ time points as independent observations sharing the same weights, whereas information about the imagined content might not be uniformly distributed across them. Let $\mathbf{W} \in \mathbb{R}^{T \times T}$ be a learnable matrix; the input is remapped as:
$$\tilde{X}_{(b,v,t)} = \sum_{s=1}^{T} W_{t,s} X_{b,v,s}, \qquad \tilde{\mathbf{X}} \in \mathbb{R}^{B \times V \times T},$$
so that every output time point is a learned linear combination of the $T$ input ones, with the same weights for every voxel. The operation is implemented as a one-dimensional convolution with $T$ input channels, $T$ output channels and kernel size $1$ along the voxel axis, without bias: a wider kernel would not be meaningful, since voxels that are adjacent in the flattened vector are not adjacent in the brain. The matrix is initialized to the identity, $\mathbf{W}^{(0)} = \mathbf{I}_T$, so that at step $0$ the model coincides exactly with the one without temporal mixing.\\
The second transformation can be represented as a multidimensional function that maps the mixed voxel data from a dimensional space $V$ to a vector $\mathbf{h} = \mathbf{f}(\tilde{\mathbf{X}}; \theta)$ of dimension $L = 8000$, applied identically to each of the $T$ time points. Next, we performed standard layer normalization, rescaling $\mathbf{h}$ with its mean value and variance, obtaining $\hat{\mathbf{h}}$. Then we applied the non linear transformation $\mathbf{z} = \mathrm{GELU}(\hat{\mathbf{h}})$, followed by dropout with probability $p = 0.25$, and projected the result into a $V_{vis}$ dimensional space, as expected from the pretrained model, by a function $\mathbf{g}(\mathbf{z};\omega)$.\\
Then we used the resulting tensor as input for the rest of the DynaDiff backbone architecture modules, which we collectively indicate as a function that maps time-series neural data from the visual cortex to the corresponding space of the generated images of $3$ channels and size $512 \times 512$ pixel: $\mathcal{D}: \mathbb{R}^{B \times V_{vis} \times T} \rightarrow \mathbb{R}^{B \times 3 \times 512 \times 512}$.
\begin{table*}[ht]
\centering
\small
\renewcommand{\arraystretch}{1.1}

\caption{Number of voxels in different ROIs for each subject.}
\label{rois_voxels}
\tablebox{
\resizebox{0.92\textwidth}{!}{%
\begin{tabular}{cccccc}
\toprule
\textbf{Subject} & \textbf{nsdgeneral} & \textbf{prefrontal cortex} &
\textbf{frontal cortex} & \textbf{temporal lobes} & \textbf{parietal cortex} \\
\midrule
$1$ & $15724$ & $14969$ & $10553$ & $12166$ & $14200$ \\
$2$ & $14278$ & $15995$ & $11360$ & $12487$ & $12494$ \\
$5$ & $13039$ & $12293$ &  $9792$ & $10811$ & $11408$ \\
$7$ & $12682$ & $12107$ &  $8760$ & $10240$ & $10430$ \\
\bottomrule
\end{tabular}%
}
}
\end{table*}
\section{Ablation Studies}\label{app_ablation}
\begin{figure}[h!]
\centering
\includegraphics[width=\linewidth]{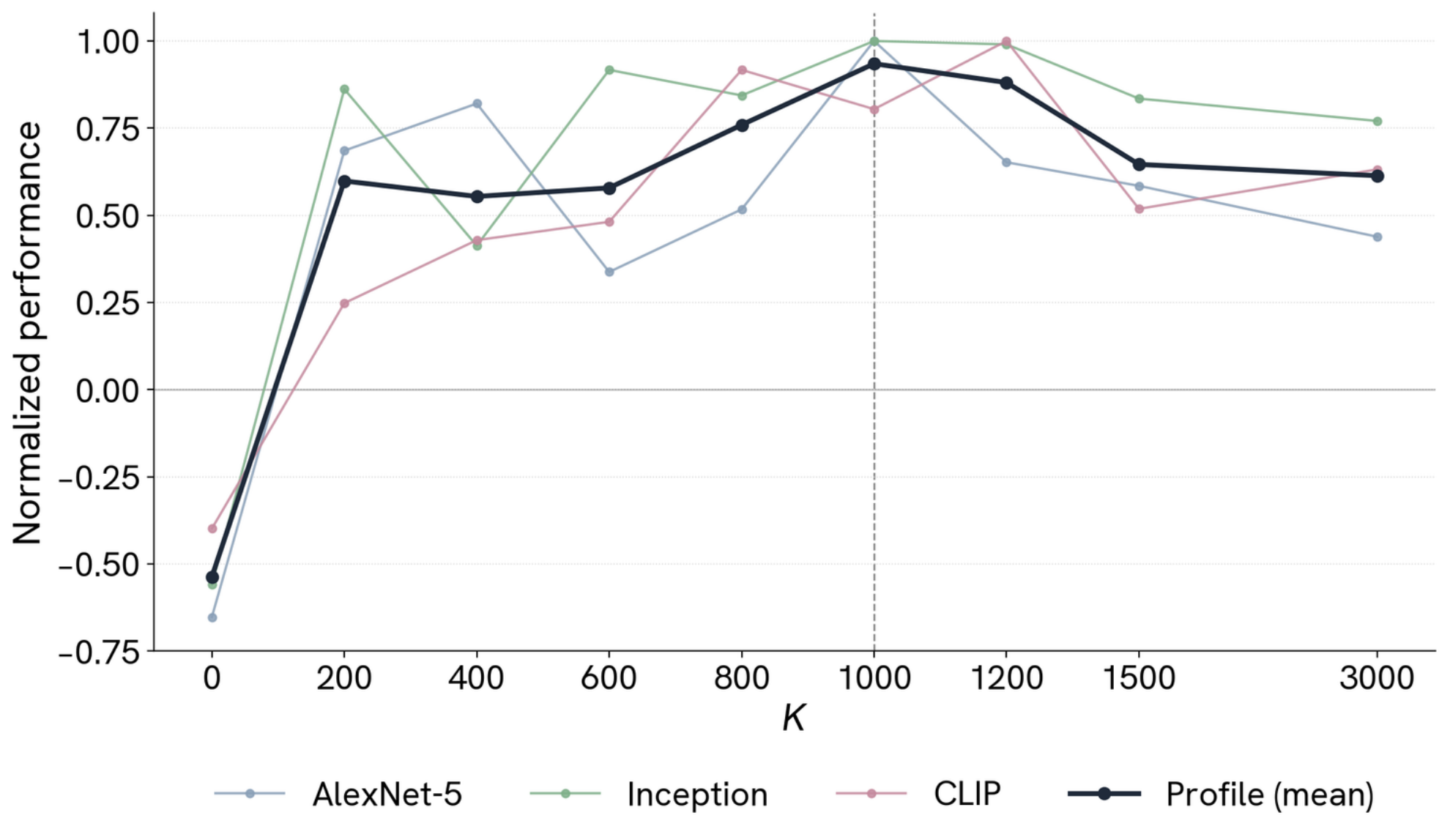}
\caption{Performance profiles of the three high-level semantic metrics obtained with LFA relative to the DynaDiff baseline, together with their mean (black line). Values represent the normalized variation in performance across the evaluated values of $K$.}
\label{ablation_k}
\end{figure}

We conducted ablation experiments to assess the individual and combined contributions of our two main components: data augmentation and LFA. All experiments were conducted on Subject 1 and evaluated on complex stimuli. We first evaluated the effect of varying the number of selected NSD samples, $K$, using our proposed LFA approach. In particular, we considered the three high-level semantic metrics AlexNet-5, CLIP, and Inception. Fig. \ref{ablation_k} shows the performance profile of each metric across the evaluated values of $K$, together with their mean profile (black line).
For each metric $m$ and value of $K$, we measured the relative change with respect to the DynaDiff baseline, $\Delta_{K,m}$, and normalized it by the maximum variation observed for that metric:
\begin{equation} 
s_{K,m}=\frac{\Delta_{K,m}}{\max_{K'}(\Delta_{K',m})}.
\end{equation}
Here, $s_{K,m}$ denotes the normalized score of metric $m$ at the selected value of $K$. Based on the mean profile across the three semantic metrics, we selected $K=1000$ as the reference value for our augmentation strategy. This value corresponds to the maximum of the mean profile, after which performance starts to decrease.

After choosing the value for $K$ we conducted two further analysis to quantify the impact of both the augmentation and our Alignment module. From Tab. \ref{tab:ablation-2x2} panel (a) we can see that augmentation without LFA, when  performing functional alignment in the voxel space, still provides an increase in the metrics, while their combination provides the best results. 
Panel (b) further investigates the contribution of non-linearity to the Alignment module by comparing a linear mapping in the CLIP embedding space with increasingly expressive architectures. Adding the convolutional component consistently improves performance, highlighting the importance of modeling the temporal dimension of the fMRI data. Introducing non-linearity in the voxel-space mapping provides a further improvement, resulting in the best overall performance.

\begin{table}[H]
\centering
\small
\renewcommand{\arraystretch}{1.1}

\caption{Ablation study of augmentation and Alignment module architecture on Subject 1 for complex stimuli with $K=1000$ and 10 partitions. Values are reported as mean $\pm$ standard deviation across partitions. Panel (a) evaluates augmentation and the Alignment module in a $2\times2$ design, while panel (b) investigates the role of non-linearity in the Alignment module. Arrows indicate whether higher ($\uparrow$) or lower ($\downarrow$) values correspond to better performance.}
\label{tab:ablation-2x2}
\tablebox{
\resizebox{0.92\textwidth}{!}{%
\begin{tabular}{llccc}
\toprule
\multicolumn{5}{c}{\textbf{(a) Augmentation $\times$ LFA}} \\
\midrule
\textbf{Augmentation} & \textbf{LFA} &
\multicolumn{3}{c}{\textbf{High-Level}} \\
\cmidrule(lr){3-5}
& & AlexNet-5 $\uparrow$ & Incep $\uparrow$ & CLIP $\uparrow$ \\
\midrule
No  & No  & $55.30 \pm 5.77\%$  & $66.97 \pm 15.43\%$ & $55.76 \pm 7.17\%$ \\
Yes & No  & $67.73 \pm 11.52\%$ & $71.06 \pm 8.31\%$  & $64.24 \pm 7.08\%$ \\
No  & Yes & $46.52 \pm 12.80\%$ & $57.73 \pm 12.65\%$ & $47.73 \pm 13.37\%$ \\
Yes & Yes & $\mathbf{68.79 \pm 12.08\%}$ & $\mathbf{83.48 \pm 8.03\%}$ & $\mathbf{71.97 \pm 5.77\%}$ \\
\midrule
\multicolumn{5}{c}{\textbf{(b) Linearity vs Non-Linearity}} \\
\midrule
\textbf{Augmentation} & \textbf{Architecture} &
\multicolumn{3}{c}{\textbf{High-Level}} \\
\cmidrule(lr){3-5}
& & AlexNet-5 $\uparrow$ & Incep $\uparrow$ & CLIP $\uparrow$ \\
\midrule
Yes & Linear             & $54.09 \pm 12.35\%$ & $75.15 \pm 8.02\%$ & $60.00 \pm 11.70\%$ \\
Yes & Linear + Conv       & $60.15 \pm 10.00\%$ & $81.67 \pm 6.52\%$ & $61.52 \pm 10.26\%$ \\
Yes & Non-Linear + Conv   & $\mathbf{68.79 \pm 12.08\%}$ & $\mathbf{83.48 \pm 8.03\%}$ & $\mathbf{71.97 \pm 5.77\%}$ \\
\bottomrule
\end{tabular}%
}
}
\end{table}

\section{Statistical Tests}\label{stats}

\begin{table}[H]
\centering
\small
\renewcommand{\arraystretch}{1.1}

\caption{Subject-wise statistical comparison, for complex stimuli, between DynaDiff Baseline and LFA for the AlexNet-5, CLIP and Inception metrics, from visual cortex. Significance was assessed using a one-sided Wilcoxon signed-rank test on paired samples. Effect sizes are reported as the median of paired differences (Median $\Delta$) with 95\% bootstrap confidence intervals. Bold marks statistically significant gains ($p<0.05$).}
\label{tab:appendix_stats}
\tablebox{
\resizebox{0.92\textwidth}{!}{%
\begin{tabular}{llcccccc}
\toprule
Metric & Subj & DynaDiff & LFA & Median $\Delta$ & 95\% CI & $W$ & $p$ \\
\midrule
AlexNet-5 & 1 & $55.30 \pm 5.77$  & $\mathbf{68.79 \pm 12.08}$ & $+15.91$ & $[+4.55, +22.73]$  & 52 & $0.004$   \\
          & 2 & $61.21 \pm 11.91$ & $\mathbf{71.36 \pm 7.57}$   & $+11.36$ & $[-3.03, +22.73]$  & 45 & $0.042$   \\
          & 5 & $50.91 \pm 15.73$ & $\mathbf{78.03 \pm 8.09}$   & $+25.76$ & $[+19.70, +37.12]$ & 55 & $<0.001$  \\
          & 7 & $53.18 \pm 8.70$  & $50.61 \pm 17.01$  & $+0.00$  & $[-18.18, +12.12]$ & 24 & $0.624$   \\
\midrule
CLIP      & 1 & $55.76 \pm 7.17$  & $\mathbf{71.97 \pm 5.77}$   & $+18.18$ & $[+7.58, +21.21]$  & 55 & $<0.001$  \\
          & 2 & $56.97 \pm 13.56$ & $\mathbf{76.82 \pm 9.92}$   & $+19.70$ & $[-3.03, +43.94]$  & 49 & $0.014$   \\
          & 5 & $45.91 \pm 4.79$  & $\mathbf{69.85 \pm 10.96}$  & $+23.48$ & $[+15.15, +28.79]$ & 55 & $<0.001$  \\
          & 7 & $54.70 \pm 8.79$  & $51.06 \pm 10.00$  & $-2.27$  & $[-13.64, +6.06]$  & 17 & $0.862$   \\
\midrule
Inception & 1 & $66.97 \pm 15.43$ & $\mathbf{83.48 \pm 8.03}$   & $+15.15$ & $[+7.58, +24.24]$  & 52 & $0.005$   \\
          & 2 & $73.33 \pm 6.23$  & $\mathbf{82.27 \pm 7.00}$   & $+13.64$ & $[+1.52, +18.18]$  & 46 & $0.030$   \\
          & 5 & $61.67 \pm 10.45$ & $\mathbf{85.76 \pm 8.61}$   & $+23.48$ & $[+16.67, +34.85]$ & 55 & $<0.001$  \\
          & 7 & $66.06 \pm 17.25$ & $63.79 \pm 10.77$  & $-5.30$  & $[-18.18, +13.64]$ & 25 & $0.615$   \\
\bottomrule
\end{tabular}%
}
}
\end{table}

\begin{table}[H]
\centering
\small
\renewcommand{\arraystretch}{1.1}

\caption{Subject-wise statistical comparison between DynaDiff Baseline and LFA on conceptual stimuli. Significance was assessed using a one-sided Wilcoxon signed-rank test on paired samples. Effect sizes are reported as the median of paired differences (Median $\Delta$) with 95\% bootstrap confidence intervals. Bold marks statistically significant gains ($p<0.05$). Baseline and LFA values are mean $\pm$ SD; the Mean row is the group mean $\pm$ SEM with a Wilcoxon test across the four subjects.}
\label{tab:appendix_stats_concept}
\tablebox{
\resizebox{0.92\textwidth}{!}{%
\begin{tabular}{lcccccc}
\toprule
Subj & Baseline & LFA & Median $\Delta$ & 95\% CI & $W$ & $p$ \\
\midrule
1 & $0.109 \pm 0.050$ & $\mathbf{0.187 \pm 0.027}$ & $+0.087$ & $[+0.043, +0.123]$ & 54 & $0.002$   \\
2 & $0.119 \pm 0.048$ & $0.156 \pm 0.041$ & $+0.068$ & $[-0.010, +0.082]$ & 40 & $0.116$   \\
5 & $0.082 \pm 0.036$ & $\mathbf{0.190 \pm 0.042}$ & $+0.082$ & $[+0.064, +0.148]$ & 55 & $<0.001$  \\
7 & $0.086 \pm 0.028$ & $\mathbf{0.152 \pm 0.035}$ & $+0.072$ & $[+0.035, +0.088]$ & 53 & $0.003$   \\
\bottomrule
\end{tabular}%
}
}
\end{table}

\begin{table}[H]
\centering
\small
\renewcommand{\arraystretch}{1.1}

\caption{Per-ROI gains of LFA over the DynaDiff baseline for complex stimuli across subjects. For each condition we report the metric improvement $\Delta$ and the paired $p$-value. Bold marks statistically significant gains ($p<0.05$). Arrows indicate whether higher ($\uparrow$) values correspond to better performance.}
\label{tab:roi-gains}
\tablebox{
\resizebox{0.92\textwidth}{!}{%
\begin{tabular}{lcccccc}
\toprule
\textbf{ROI} &
\multicolumn{6}{c}{\textbf{High-Level metrics (Complex stimuli)}} \\
\cmidrule(lr){2-7}
&
\multicolumn{2}{c}{AlexNet-5 $\uparrow$} &
\multicolumn{2}{c}{Incep $\uparrow$} &
\multicolumn{2}{c}{CLIP $\uparrow$} \\
\cmidrule(lr){2-3} \cmidrule(lr){4-5} \cmidrule(lr){6-7}
& $\Delta$ & $p$ & $\Delta$ & $p$ & $\Delta$ & $p$ \\
\midrule
\multicolumn{7}{c}{\textbf{Single ROI}} \\
\midrule
Parietal    & $\mathbf{+8.03}$ & $0.005$ & $\mathbf{+10.68}$ & $0.001$ & $\mathbf{+14.96}$ & $0.001$ \\
Frontal     & $+3.64$          & $0.116$ & $\mathbf{+5.83}$  & $0.032$ & $+3.94$           & $0.080$ \\
Temporal    & $\mathbf{+6.52}$ & $0.002$ & $\mathbf{+4.43}$  & $0.042$ & $\mathbf{+5.38}$  & $0.005$ \\
Prefrontal  & $+0.91$          & $0.385$ & $+1.17$           & $0.216$ & $+0.27$           & $0.278$ \\
\midrule
\multicolumn{7}{c}{\textbf{ROI $\cup$ nsdgeneral}} \\
\midrule
Parietal    & $\mathbf{+13.64}$ & $0.001$ & $\mathbf{+13.26}$ & $0.001$ & $\mathbf{+12.95}$ & $0.001$ \\
Frontal     & $\mathbf{+11.67}$ & $0.001$ & $\mathbf{+8.30}$  & $0.003$ & $\mathbf{+10.38}$ & $0.003$ \\
Temporal    & $\mathbf{+11.48}$ & $0.001$ & $\mathbf{+9.89}$  & $0.001$ & $\mathbf{+11.29}$ & $0.001$ \\
Prefrontal  & $\mathbf{+7.35}$  & $0.010$ & $\mathbf{+8.45}$  & $0.002$ & $\mathbf{+9.85}$  & $0.003$ \\
\bottomrule
\end{tabular}%
}
}
\end{table}

\begin{table}[H]
\centering
\small
\renewcommand{\arraystretch}{1.1}

\caption{Per-ROI similarity gains of LFA over the DynaDiff Baseline for conceptual stimuli across subjects. For each ROI we report the improvement $\Delta$ and the paired $p$-value, both for the ROI alone and for the ROI combined with \textit{nsdgeneral}. Bold marks statistically significant gains ($p<0.05$). Arrows indicate whether higher ($\uparrow$) values correspond to better performance.}
\label{tab:roi-similarity}
\tablebox{
\resizebox{0.92\textwidth}{!}{%
\begin{tabular}{lcccc}
\toprule &
\multicolumn{2}{c}{\textbf{Single ROI}} &
\multicolumn{2}{c}{\textbf{ROI $\cup$ nsdgeneral}} \\
\cmidrule(lr){2-3} \cmidrule(lr){4-5}
& $\Delta$ Similarity $\uparrow$ & $p$
& $\Delta$ Similarity $\uparrow$ & $p$ \\
\midrule
\multicolumn{5}{c}{\textbf{Conceptual stimuli}} \\
\midrule
Parietal &
 $\mathbf{+0.055}$ & $0.001$ &
 $\mathbf{+0.073}$ & $0.001$ \\
Frontal &
 $\mathbf{+0.049}$ & $0.001$ &
 $\mathbf{+0.064}$ & $0.001$ \\
Temporal &
 $\mathbf{+0.070}$ & $0.001$ &
 $\mathbf{+0.073}$ & $0.001$ \\
Prefrontal &
 $\mathbf{+0.051}$ & $0.001$ &
 $\mathbf{+0.060}$ & $0.001$ \\
\midrule
\multicolumn{5}{c}{\textbf{Reference: nsdgeneral (LFA)} \quad $\Delta = \mathbf{+0.072}$, \; $p = 0.001$} \\
\bottomrule
\end{tabular}%
}
}
\end{table}

\section{Quantitative Results}\label{app_quantitative}
\begin{table}[H]
\centering
\small
\renewcommand{\arraystretch}{1.1}

\caption{Reconstruction of NSD-Imagery \emph{vision} trials, decoded by the pretrained model without any alignment. All trials of a stimulus are averaged into a single input, so one reconstruction per stimulus is produced. Values are mean $\pm$ SEM across the four subjects. Arrows indicate whether higher ($\uparrow$) or lower ($\downarrow$) values correspond to better performance.}
\label{tab:vision_mean}
\tablebox{
\resizebox{0.92\textwidth}{!}{%
\begin{tabular}{lcccccccc}
\toprule
\textbf{Stimuli} &
\multicolumn{2}{c}{\textbf{Low-Level}} &
\multicolumn{4}{c}{\textbf{High-Level}} &
\multicolumn{2}{c}{\textbf{Distance}} \\
\cmidrule(lr){2-3} \cmidrule(lr){4-7} \cmidrule(lr){8-9}
& PixCorr $\uparrow$ & SSIM $\uparrow$
& AlexNet-2 $\uparrow$ & AlexNet-5 $\uparrow$ & Incep $\uparrow$ & CLIP $\uparrow$
& Eff $\downarrow$ & SwAV $\downarrow$ \\
\midrule
\multicolumn{9}{c}{\textbf{Vision-task -- All Subjects -- Visual Cortex}} \\
\midrule
Simple stimuli & $0.094 \pm 0.021$ & $0.548 \pm 0.027$ & $66.67 \pm 6.28\%$ & $81.82 \pm 3.76\%$ & $50.00 \pm 5.64\%$ & $62.50 \pm 5.91\%$ & $0.986 \pm 0.013$ & $0.561 \pm 0.011$ \\
Complex stimuli & $0.137 \pm 0.011$ & $0.284 \pm 0.005$ & $91.67 \pm 0.76\%$ & $87.12 \pm 3.06\%$ & $80.68 \pm 5.51\%$ & $87.88 \pm 3.86\%$ & $0.874 \pm 0.013$ & $0.511 \pm 0.003$ \\
\bottomrule
\end{tabular}%
}
}
\end{table}

\begin{table}[H]
\centering
\small
\renewcommand{\arraystretch}{1.1}

\caption{Per-subject results (mean $\pm$ SD over the 10 data partitions) and average across subjects (mean $\pm$ SEM across the four subjects) for imagined simple and complex stimuli reconstruction from visual cortex. Since no alignment is applied to simple stimuli, all differences originate from complex stimuli and are halved by the pooling. Arrows indicate whether higher ($\uparrow$) or lower ($\downarrow$) values correspond to better performance.}
\label{tab:taskAB_persubj}
\tablebox{
\resizebox{0.92\textwidth}{!}{%
\begin{tabular}{llcccccccc}
\toprule
\textbf{Method} & \textbf{Subj} &
\multicolumn{2}{c}{\textbf{Low-Level}} &
\multicolumn{4}{c}{\textbf{High-Level}} &
\multicolumn{2}{c}{\textbf{Distance}} \\
\cmidrule(lr){3-4} \cmidrule(lr){5-8} \cmidrule(lr){9-10}
& & PixCorr $\uparrow$ & SSIM $\uparrow$
& AlexNet-2 $\uparrow$ & AlexNet-5 $\uparrow$ & Incep $\uparrow$ & CLIP $\uparrow$
& Eff $\downarrow$ & SwAV $\downarrow$ \\
\midrule
\multicolumn{10}{c}{\textbf{Simple and Complex stimuli -- Imagination-task -- Visual Cortex}} \\
\midrule
\multirow{5}{*}{\makecell[l]{DynaDiff\\Baseline}}
 & 1 & $0.019 \pm 0.029$ & $0.373 \pm 0.023$ & $64.62 \pm 8.38\%$ & $62.35 \pm 5.22\%$ & $57.42 \pm 7.98\%$ & $59.09 \pm 4.82\%$ & $0.976 \pm 0.017$ & $0.602 \pm 0.016$ \\
 & 2 & $-0.011 \pm 0.031$ & $0.360 \pm 0.016$ & $58.18 \pm 7.20\%$ & $62.58 \pm 8.92\%$ & $62.12 \pm 4.32\%$ & $57.35 \pm 5.88\%$ & $0.978 \pm 0.022$ & $0.600 \pm 0.013$ \\
 & 5 & $-0.018 \pm 0.040$ & $0.354 \pm 0.038$ & $54.32 \pm 7.84\%$ & $53.71 \pm 12.47\%$ & $52.58 \pm 5.78\%$ & $54.70 \pm 4.18\%$ & $0.977 \pm 0.013$ & $0.608 \pm 0.017$ \\
 & 7 & $0.002 \pm 0.025$ & $0.361 \pm 0.026$ & $53.71 \pm 7.41\%$ & $60.30 \pm 5.47\%$ & $52.65 \pm 9.96\%$ & $59.92 \pm 4.73\%$ & $0.988 \pm 0.017$ & $0.606 \pm 0.016$ \\
\cmidrule(lr){2-10}
 & Mean & $-0.002 \pm 0.008$ & $0.362 \pm 0.004$ & $57.71 \pm 2.51\%$ & $59.73 \pm 2.07\%$ & $56.19 \pm 2.28\%$ & $57.77 \pm 1.15\%$ & $0.979 \pm 0.003$ & $0.604 \pm 0.002$ \\
\midrule
\multirow{5}{*}{LFA}
 & 1 & $0.034 \pm 0.030$ & $0.382 \pm 0.028$ & $65.45 \pm 4.90\%$ & $69.09 \pm 7.81\%$ & $65.68 \pm 6.09\%$ & $67.20 \pm 5.48\%$ & $0.949 \pm 0.024$ & $0.580 \pm 0.024$ \\
 & 2 & $0.019 \pm 0.038$ & $0.367 \pm 0.017$ & $63.03 \pm 5.10\%$ & $67.65 \pm 3.44\%$ & $66.59 \pm 6.84\%$ & $67.27 \pm 6.56\%$ & $0.938 \pm 0.023$ & $0.570 \pm 0.019$ \\
 & 5 & $0.004 \pm 0.034$ & $0.358 \pm 0.041$ & $58.71 \pm 7.29\%$ & $67.27 \pm 8.08\%$ & $64.62 \pm 4.86\%$ & $66.67 \pm 7.40\%$ & $0.940 \pm 0.023$ & $0.582 \pm 0.016$ \\
 & 7 & $0.014 \pm 0.029$ & $0.358 \pm 0.037$ & $51.89 \pm 4.38\%$ & $59.02 \pm 9.46\%$ & $51.52 \pm 8.44\%$ & $58.11 \pm 5.50\%$ & $0.985 \pm 0.015$ & $0.607 \pm 0.022$ \\
\cmidrule(lr){2-10}
 & Mean & $\mathbf{0.018 \pm 0.006}$ & $\mathbf{0.366 \pm 0.006}$ & $\mathbf{59.77 \pm 2.97\%}$ & $\mathbf{65.76 \pm 2.28\%}$ & $\mathbf{62.10 \pm 3.55\%}$ & $\mathbf{64.81 \pm 2.24\%}$ & $\mathbf{0.953 \pm 0.011}$ & $\mathbf{0.585 \pm 0.008}$ \\
\bottomrule
\end{tabular}%
}
}
\end{table}

\begin{table}[H]
\centering
\small
\renewcommand{\arraystretch}{1.1}

\caption{Per-subject results (mean $\pm$ SD over the 10 data partitions) and average across subjects (mean $\pm$ SEM across the four subjects) for imagined simple stimuli reconstructions. No alignment is applied to simple stimuli, so the DynaDiff baseline and LFA coincide by construction. Arrows indicate whether higher ($\uparrow$) or lower ($\downarrow$) values correspond to better performance.}
\label{tab:imagery_simple_persubj}
\tablebox{
\resizebox{0.92\textwidth}{!}{%
\begin{tabular}{lcccccccc}
\toprule
\textbf{Subj} &
\multicolumn{2}{c}{\textbf{Low-Level}} &
\multicolumn{4}{c}{\textbf{High-Level}} &
\multicolumn{2}{c}{\textbf{Distance}} \\
\cmidrule(lr){2-3} \cmidrule(lr){4-7} \cmidrule(lr){8-9}
& PixCorr $\uparrow$ & SSIM $\uparrow$
& AlexNet-2 $\uparrow$ & AlexNet-5 $\uparrow$ & Incep $\uparrow$ & CLIP $\uparrow$
& Eff $\downarrow$ & SwAV $\downarrow$ \\
\midrule
\multicolumn{9}{c}{\textbf{Simple stimuli -- Imagination-task -- Visual Cortex}} \\
\midrule
1 & $0.020 \pm 0.023$ & $0.498 \pm 0.053$ & $49.55 \pm 14.73\%$ & $69.39 \pm 6.05\%$ & $47.88 \pm 10.06\%$ & $62.42 \pm 6.99\%$ & $0.988 \pm 0.023$ & $0.581 \pm 0.015$ \\
2 & $-0.007 \pm 0.036$ & $0.476 \pm 0.024$ & $36.97 \pm 10.69\%$ & $63.94 \pm 8.26\%$ & $50.91 \pm 7.77\%$ & $57.73 \pm 5.07\%$ & $1.001 \pm 0.018$ & $0.586 \pm 0.021$ \\
5 & $-0.012 \pm 0.040$ & $0.474 \pm 0.079$ & $33.18 \pm 11.46\%$ & $56.52 \pm 16.00\%$ & $43.48 \pm 9.23\%$ & $63.48 \pm 8.61\%$ & $0.997 \pm 0.013$ & $0.598 \pm 0.026$ \\
7 & $-0.001 \pm 0.024$ & $0.485 \pm 0.059$ & $34.55 \pm 8.92\%$ & $67.42 \pm 9.46\%$ & $39.24 \pm 10.87\%$ & $65.15 \pm 4.74\%$ & $1.005 \pm 0.016$ & $0.582 \pm 0.029$ \\
\cmidrule(lr){1-9}
Mean & $\mathbf{0.000 \pm 0.007}$ & $\mathbf{0.483 \pm 0.005}$ & $\mathbf{38.56 \pm 3.74\%}$ & $\mathbf{64.32 \pm 2.83\%}$ & $\mathbf{45.38 \pm 2.55\%}$ & $\mathbf{62.20 \pm 1.59\%}$ & $\mathbf{0.998 \pm 0.004}$ & $\mathbf{0.587 \pm 0.004}$ \\
\bottomrule
\end{tabular}%
}
}
\end{table}

\begin{table}[H]
\centering
\small
\renewcommand{\arraystretch}{1.1}

\caption{Per-subject results (mean $\pm$ SD over the 10 data partitions) and average across subjects (mean $\pm$ SEM across the four subjects) for imagined complex stimuli reconstructions. Arrows indicate whether higher ($\uparrow$) or lower ($\downarrow$) values correspond to better performance.}
\label{tab:taskB_persubj}
\tablebox{
\resizebox{0.92\textwidth}{!}{%
\begin{tabular}{llcccccccc}
\toprule
\textbf{Method} & \textbf{Subj} &
\multicolumn{2}{c}{\textbf{Low-Level}} &
\multicolumn{4}{c}{\textbf{High-Level}} &
\multicolumn{2}{c}{\textbf{Distance}} \\
\cmidrule(lr){3-4} \cmidrule(lr){5-8} \cmidrule(lr){9-10}
& & PixCorr $\uparrow$ & SSIM $\uparrow$
& AlexNet-2 $\uparrow$ & AlexNet-5 $\uparrow$ & Incep $\uparrow$ & CLIP $\uparrow$
& Eff $\downarrow$ & SwAV $\downarrow$ \\
\midrule
\multicolumn{10}{c}{\textbf{Complex stimuli -- Imagination-task -- Visual Cortex}} \\
\midrule
\multirow{5}{*}{\makecell[l]{DynaDiff\\Baseline}}
 & 1 & $0.018 \pm 0.051$ & $0.248 \pm 0.017$ & $79.70 \pm 5.90\%$ & $55.30 \pm 5.77\%$ & $66.97 \pm 15.43\%$ & $55.76 \pm 7.17\%$ & $0.963 \pm 0.032$ & $0.623 \pm 0.024$ \\
 & 2 & $-0.014 \pm 0.060$ & $0.245 \pm 0.026$ & $79.39 \pm 5.45\%$ & $61.21 \pm 11.91\%$ & $73.33 \pm 6.23\%$ & $56.97 \pm 13.56\%$ & $0.954 \pm 0.033$ & $0.615 \pm 0.017$ \\
 & 5 & $-0.025 \pm 0.074$ & $0.233 \pm 0.027$ & $75.45 \pm 8.32\%$ & $50.91 \pm 15.73\%$ & $61.67 \pm 10.45\%$ & $45.91 \pm 4.79\%$ & $0.956 \pm 0.026$ & $0.618 \pm 0.020$ \\
 & 7 & $0.006 \pm 0.033$ & $0.237 \pm 0.024$ & $72.88 \pm 9.83\%$ & $53.18 \pm 8.70\%$ & $66.06 \pm 17.25\%$ & $54.70 \pm 8.79\%$ & $0.971 \pm 0.022$ & $0.631 \pm 0.023$ \\
\cmidrule(lr){2-10}
 & Mean & $-0.004 \pm 0.010$ & $0.241 \pm 0.003$ & $76.86 \pm 1.64\%$ & $55.15 \pm 2.21\%$ & $67.01 \pm 2.41\%$ & $53.33 \pm 2.52\%$ & $0.961 \pm 0.004$ & $0.622 \pm 0.003$ \\
\midrule
\multirow{5}{*}{LFA}
 & 1 & $0.048 \pm 0.059$ & $0.265 \pm 0.020$ & $81.36 \pm 9.20\%$ & $68.79 \pm 12.08\%$ & $83.48 \pm 8.03\%$ & $71.97 \pm 5.77\%$ & $0.909 \pm 0.037$ & $0.580 \pm 0.043$ \\
 & 2 & $0.045 \pm 0.062$ & $0.258 \pm 0.018$ & $89.09 \pm 5.38\%$ & $71.36 \pm 7.57\%$ & $82.27 \pm 7.00\%$ & $76.82 \pm 9.92\%$ & $0.875 \pm 0.050$ & $0.555 \pm 0.038$ \\
 & 5 & $0.021 \pm 0.054$ & $0.241 \pm 0.020$ & $84.24 \pm 8.21\%$ & $78.03 \pm 8.09\%$ & $85.76 \pm 8.61\%$ & $69.85 \pm 10.96\%$ & $0.884 \pm 0.046$ & $0.566 \pm 0.040$ \\
 & 7 & $0.028 \pm 0.047$ & $0.232 \pm 0.030$ & $69.24 \pm 6.43\%$ & $50.61 \pm 17.01\%$ & $63.79 \pm 10.77\%$ & $51.06 \pm 10.00\%$ & $0.964 \pm 0.026$ & $0.633 \pm 0.022$ \\
\cmidrule(lr){2-10}
 & Mean & $\mathbf{0.035 \pm 0.007}$ & $\mathbf{0.249 \pm 0.008}$ & $\mathbf{80.98 \pm 4.23\%}$ & $\mathbf{67.20 \pm 5.86\%}$ & $\mathbf{78.83 \pm 5.06\%}$ & $\mathbf{67.42 \pm 5.65\%}$ & $\mathbf{0.908 \pm 0.020}$ & $\mathbf{0.583 \pm 0.017}$ \\
\bottomrule
\end{tabular}%
}
}
\end{table}

\begin{table}[H]
\centering
\small
\renewcommand{\arraystretch}{1.1}

\caption{Comparison of Brain Diffuser, MindEye1, MindEye2, DynaDiff Baseline and our LFA method on simple and complex stimuli reconstructions (12 stimuli). Uncertainties were not reported for other SOTA models in \cite{kneeland2025nsdimagerybenchmarkdatasetextending} and we omitted them from our results, however they are reported in Tab. \ref{tab:taskAB_persubj}.}
\label{tab}
\tablebox{
\resizebox{0.92\textwidth}{!}{%
\begin{tabular}{lcccccccc}
\toprule
\textbf{Method} &
\multicolumn{2}{c}{\textbf{Low-Level}} &
\multicolumn{4}{c}{\textbf{High-Level}} &
\multicolumn{2}{c}{\textbf{Distance}} \\
\cmidrule(lr){2-3} \cmidrule(lr){4-7} \cmidrule(lr){8-9}
& PixCorr $\uparrow$ & SSIM $\uparrow$
& AlexNet-2 $\uparrow$ & AlexNet-5 $\uparrow$ & Incep $\uparrow$ & CLIP $\uparrow$
& Eff $\downarrow$ & SwAV $\downarrow$ \\
\midrule
\multicolumn{9}{c}{\textbf{Simple and Complex stimuli -- Imagination-task -- Visual Cortex}} \\
\midrule
MindEye1 &
 $\mathbf{0.086}$ &
 $0.349$ &
 $59.56\%$ &
 $61.00\%$ &
 $52.03\%$ &
 $54.72\%$ &
 $0.948$ &
 $\mathbf{0.564}$ \\
Brain Diffuser &
 $0.064$ &
 $0.401$ &
 $52.14\%$ &
 $58.35\%$ &
 $52.73\%$ &
 $54.07\%$ &
 $\mathbf{0.935}$ &
 $0.585$ \\
MindEye2 &
 $0.036$ &
 $\mathbf{0.414}$ &
 $47.60\%$ &
 $55.38\%$ &
 $46.02\%$ &
 $50.78\%$ &
 $0.966$ &
 $0.591$ \\
DynaDiff Baseline &
 $-0.002 $ &
 $0.362 $ &
 $57.71 \%$ &
 $59.73 \%$ &
 $56.19 \%$ &
 $57.77 \%$ &
 $0.979$ &
 $0.604$ \\
LFA &
 $0.018$ &
 $0.366 $ &
 $\mathbf{59.77}\%$ &
 $\mathbf{65.76}\%$ &
 $\mathbf{62.10 }\%$ &
 $\mathbf{64.81}\%$ &
 $0.953$ &
 $0.585 $ \\
\bottomrule
\end{tabular}%
}
}
\end{table}

\section{Qualitative Results}\label{app_qualitative}
\begin{figure}[H]
  \centering
  \includegraphics[width=\linewidth]{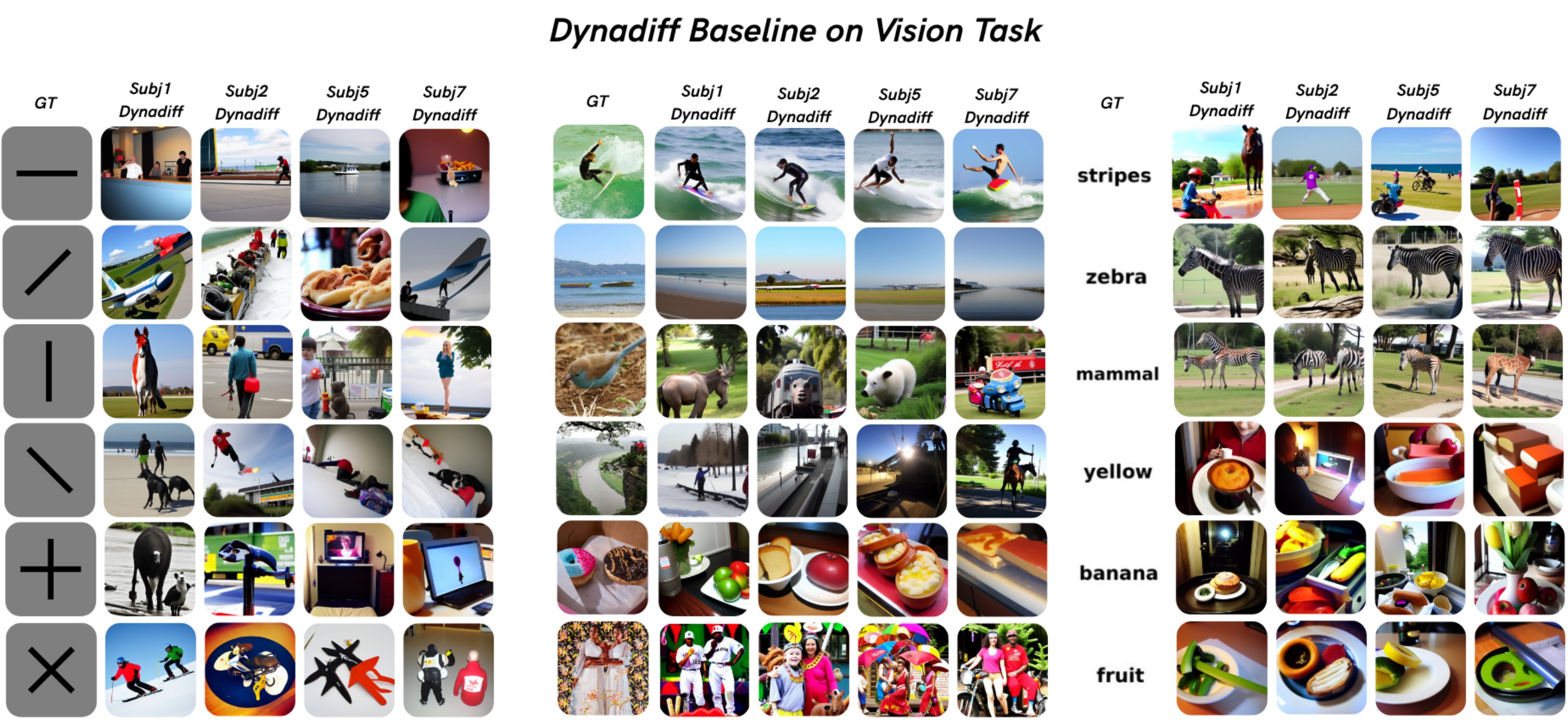}
  \caption{Qualitative results of the  reconstruction for all the subject of viewed simple, complex and conceptual stimuli, from left to right respectively.}\label{all_vision}
\end{figure}

\begin{figure}[H]
  \centering
  \includegraphics[width=\linewidth]{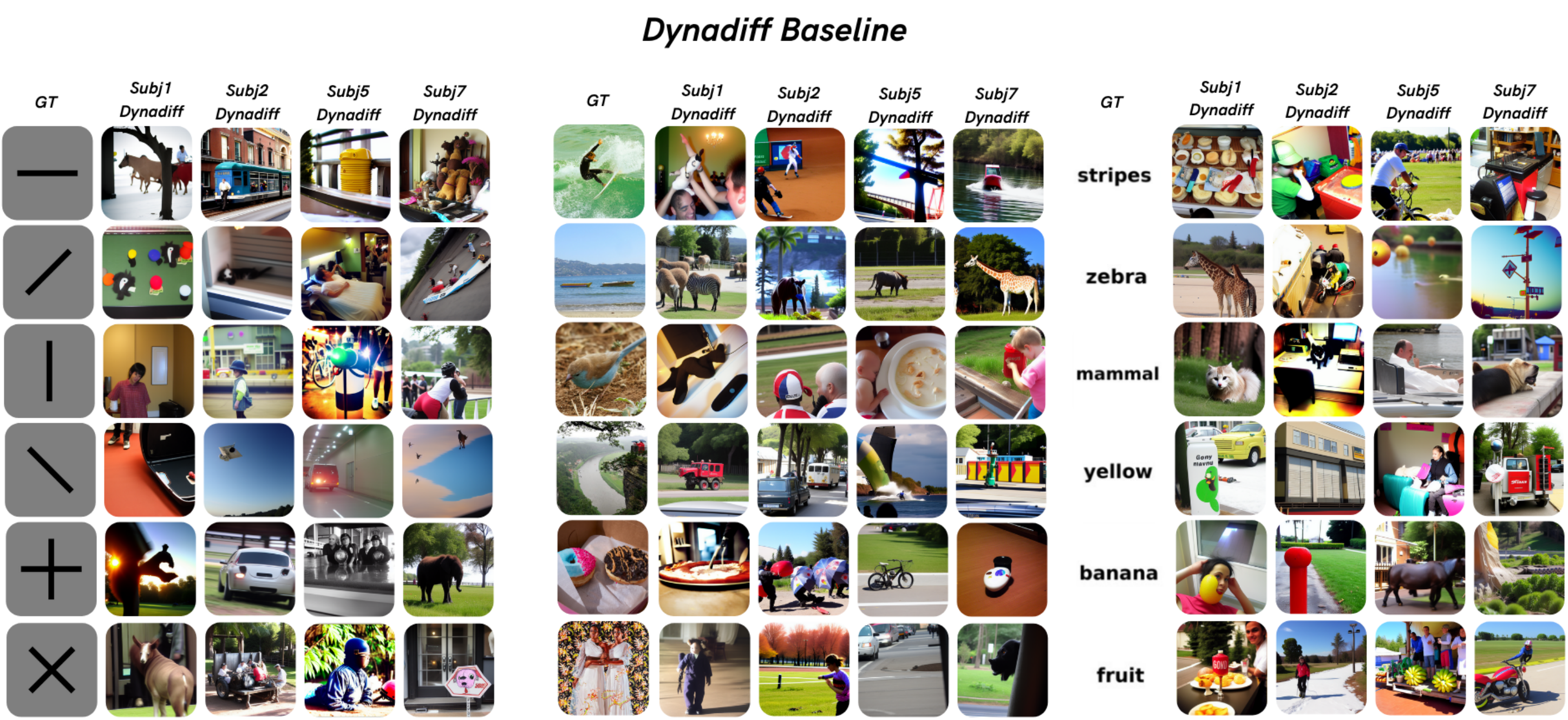}
  \caption{Qualitative results of the  reconstruction for all the subject of imagined simple, complex and conceptual stimuli, from left to right respectively.}\label{all_baseline}
\end{figure}

\begin{figure}[H]
  \centering
  \includegraphics[width=\linewidth]{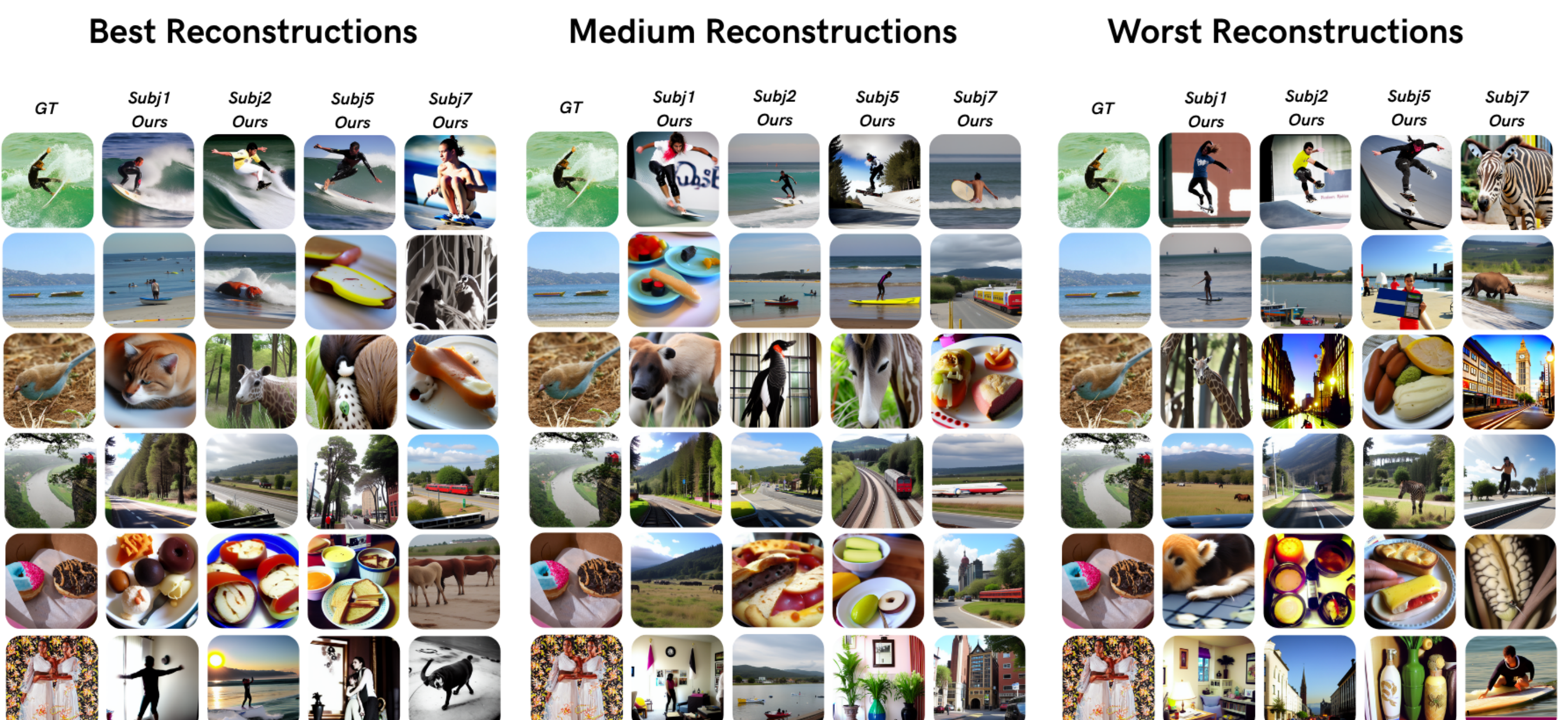}
  \caption{Qualitative results of the best, medium and worst reconstruction obtained with our LFA for all the subjects of imagined complex stimuli, from left to right respectively.}\label{all_complex}
\end{figure}

\begin{figure}[H]
  \centering
  \includegraphics[width=\linewidth]{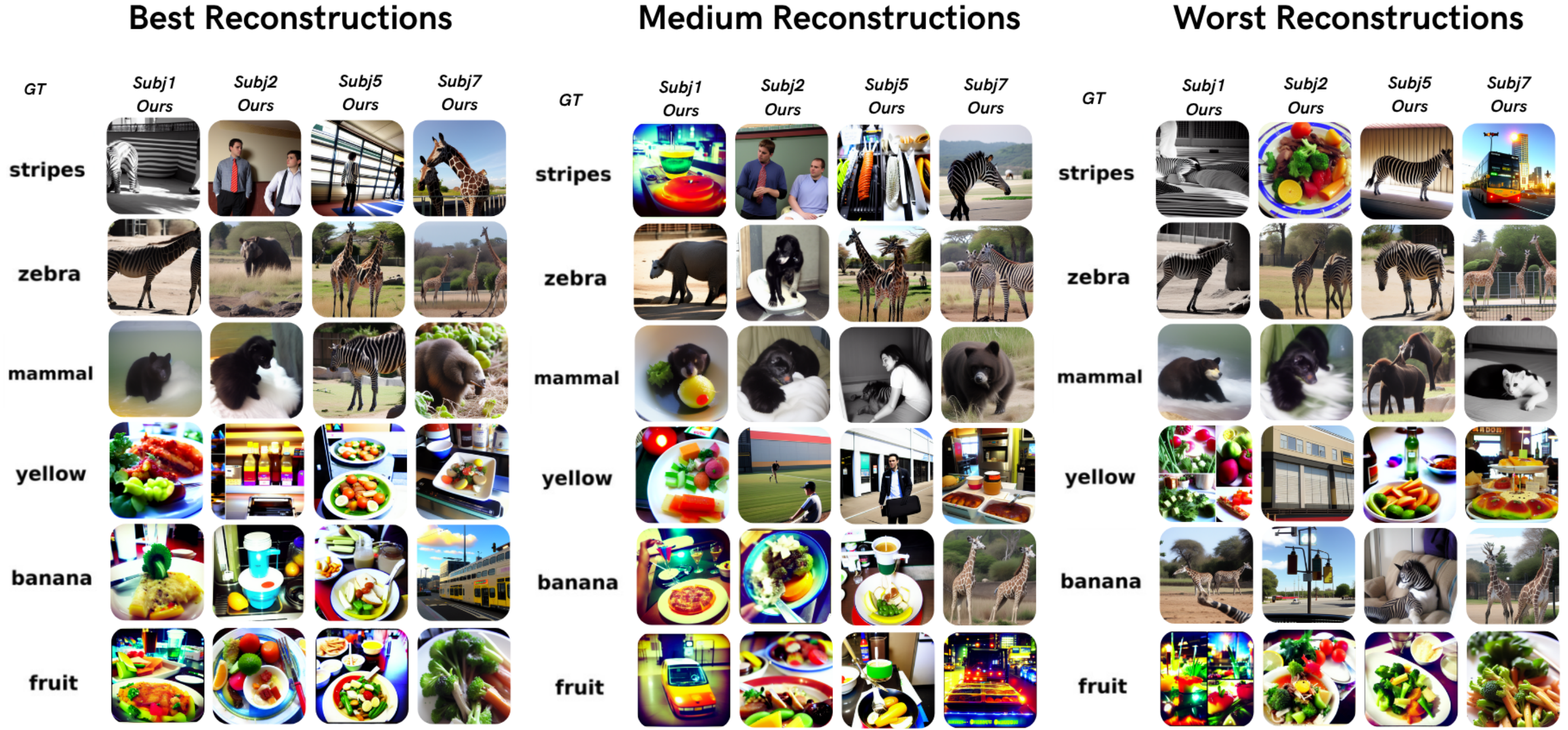}
  \caption{Qualitative results of the best, medium and worst reconstruction obtained with our LFA for all the subjects of imagined conceptual stimuli, from left to right respectively.}\label{all_concept}
\end{figure}

\end{document}